\documentclass[12pt]{article}
\usepackage{graphicx}
\usepackage{amsfonts,amsthm}
\usepackage{amsmath,amssymb}
\usepackage{bm}
\usepackage[utf8]{inputenc}
\usepackage[usenames,dvipsnames]{xcolor}

\newcommand{\be}{\begin{equation}}
\newcommand{\ee}{\end{equation}}
\newcommand{\bea}{\begin{eqnarray}}
\newcommand{\eea}{\end{eqnarray}}
\newcommand{\dst}{\displaystyle}
\newcommand{\lr}[1]{ \langle #1 \rangle}

\newcommand{\fr}[2]{\frac{{\dst #1}}{{\dst #2}}}
\renewcommand{\Re}{\mathrm{Re }}

\newcommand{\mmmatrix}[9]{ \left(\! \begin{array}{ccc}#1 & #2 & #3\\[1mm] #4 & #5 & #6\\[1mm] #7 & #8 & #9\\ \end{array}\!\right) }
\newcommand{\doublet}[2]{ \left( \begin{array}{c}#1 \\ #2 \end{array}\right) }
\newcommand{\triplet}[3]{ \left( \begin{array}{c}#1 \\[2mm] #2 \\[2mm] #3\end{array}\right) }

\newcommand{\bA}{{\bf A}}

\newcommand{\bbe}{{\bf e}}
\newcommand{\bk}{{\bf k}}
\newcommand{\bK}{{\bf K}}
\newcommand{\bn}{{\bf n}}

\newcommand{\br}{{\bf r}}

\newcommand{\GeV}{\mathrm{GeV}}

\def\lsim{\mathrel{\rlap{\lower4pt\hbox{\hskip1pt$\sim$}}
    \raise1pt\hbox{$<$}}}         
\def\gsim{\mathrel{\rlap{\lower4pt\hbox{\hskip1pt$\sim$}}
    \raise1pt\hbox{$>$}}}         

    \usepackage{bbold}

\graphicspath{{./figs/}}

\newcommand{\ga}{\gamma}
\newcommand{\de}{\delta}

\newcommand{\la}{\lambda}

\newcommand{\si}{\sigma}

\title{{\normalsize \hfill CFTP/20-001} \\*[7mm]
 Twisted particle collisions: a new tool for spin physics}

\author{Igor P. Ivanov$^1$, Nikolai Korchagin$^2$, Alexandr Pimikov$^{2,3}$, Pengming Zhang$^{4}$
	\\[5mm]
	{\small $^1$ CFTP, Instituto Superior Tecnico, Universidade de Lisboa, Lisbon 1049-001, Portugal}\\
	{\small $^2$ Institute of Modern Physics, Chinese Academy of Sciences, Lanzhou 730000, China}\\
	{\small $^3$ Research Institute of Physics, Southern Federal University, Rostov-na-Donu 344090, Russia}\\
	{\small $^4$ School of Physics and Astronomy, Sun Yat-sen University, Zhuhai 519082, China}\\
}

\begin{document}

\maketitle

\begin{abstract}
Collisions of twisted particles --- that is, non-plane-wave states of photons, electrons, or any other particle,
equipped with a non-zero orbital angular momentum (OAM) with respect to its propagation direction ---
offer novel ways to probe particle structure and interactions.
In the recent paper \cite{Ivanov:2019vxe}, we argued that resonance production 
in twisted photon collisions or twisted $e^+e^-$ annihilation 
gives access to parity- and spin-sensitive observables in inclusive cross sections, 
even when the initial particles are unpolarized. 
Here, we explore these features in detail,
providing a qualitative picture and illustrating it with numerical examples.
We show how one can detect parity-violating effects in collisions of unpolarized twisted photons
and how one can produce almost $100\%$ polarized vector mesons
in unpolarized twisted $e^+e^-$ annihilation. 
These examples highlight the unprecedented level of control over polarization offered
by twisted particles, impossible in the usual plane wave collisions. 
\end{abstract}


\newpage

\section{Introduction}

Determining the spin and parity properties of hadrons
is an intricate and fascinating aspect of modern particle phenomenology.
Known hadrons, including the very short-lived resonances, exhibit a variety
of spin-parity quantum numbers possible with the usual $q\bar q$ and $qqq$
quark combinations \cite{Tanabashi:2018oca,Eichmann:2016yit}, 
with multiquark states \cite{Olsen:2014qna,Pimikov:2019dyr}
and glueball states \cite{Pimikov:2017bkk}.
Deep inelastic scattering (DIS) with polarized initial lepton or proton
allows one to investigate how spin of the ultrarelativistic proton emerges
from spins and orbital angular momenta of its constituents \cite{Aidala:2012mv}.
This problem alone, dubbed the spin proton crisis, has remained a source of controversies 
over the past decades, and the situation is still far from being completely resolved \cite{Leader:2013jra}.
Going beyond helicity distributions and entering the realm of the 3D momentum space spin structure 
brings in many new spin-sensitive variables, 
which can be encoded via transverse-momentum distributions and explored experimentally
in semi-inclusive DIS with transversely polarized protons \cite{Anselmino:2007fs,Anselmino:2020vlp}.
The recently approved Electron Ion Collider in the US \cite{Boer:2019pzo} and 
the proposed Electron Ion Collider in China \cite{Chen:2018wyz} also have a rich spin physics program ahead.

There are two main experimental tools for investigating spin-parity properties of hadrons.
First, one can produce them in collision of longitudinally or transversely polarized initial particles
and measure spin asymmetries, that is, the response of the cross section to flipping
the polarization sign or changing the polarization direction.
Second, one can study exclusive or semi-inclusive reactions,
look into angular distribution of the final state particles,
and, with the aid of partial-wave analysis, deduce the spin properties
either of the target hadron (in DIS) or of the intermediate resonances 
(in low energy exclusive production processes).
In either case, the task requires certain experimental efforts
in preparing a polarized initial state 
or in extracting the angular correlations of the final state particles.
There seems to be no other way to access spin-dependent observables.

In the recent paper \cite{Ivanov:2019vxe} we proposed a completely new tool for doing spin physics
in particle collisions. We showed that if one prepares initial particles in
the so-called twisted state, in which they are equipped with a non-zero, adjustable
orbital angular momentum (OAM) with respect to their propagation direction,
then spin- and parity-dependent observables can be probed with fully inclusive cross sections
of unpolarized particles.
In this paper, we provide a detailed exploration of this idea
by considering production of spin-0 or spin-2 resonances in twisted $\gamma\gamma$ collisions
and of spin-1 resonance in twisted $e^+e^-$ annihilation.
We present both a qualitative picture, which helps understand several consequences
which otherwise may seem counter-intuitive, and corroborate it with numerical examples.

In the following section, we give a brief reminder of how twisted photons and electrons are described and equip
the reader with expressions for calculations of the twisted helicity amplitudes. 
We also discuss in detail the subtle notion of unpolarized twisted photon or electron beam.
Then in section~\ref{section-photons} we calculate production of spin-0
and spin-2 resonances in twisted photon collisions. We show how one can extract
scalar-pseudoscalar mixing in spin-0 production and how to selectively produce
a spin-2 particle in a specific polarization state --- all done with unpolarized twisted photon beams. 
In section~\ref{section-e+e-} we repeat the analysis for vector meson
production in unpolarized twisted $e^+e^-$ annihilation. 
We summarize the results and give an outlook in the last section.

Throughout the paper, we use natural units $\hbar = c = 1$.
Three-dimensional vectors will be denoted by bold symbols,
while the transverse momenta will be labeled by the subscript~$\perp$.
We will often use the shorthand notations $c$ and $s$, which stand for:
\begin{eqnarray}
&&c \equiv \cos\theta\,, \qquad \  \ s\equiv \sin\theta\qquad\ \ \mbox{for scalar and vector fields;}\nonumber\\
&&c \equiv \cos(\theta/2)\,, \quad s\equiv \sin(\theta/2)\quad \mbox{for electrons and positrons}\,.\nonumber
\end{eqnarray}

\section{Describing twisted particles}

\subsection{Twisted scalar particles}\label{subsection-description-scalar}

Description of twisted particle states in a way convenient for 
calculations of high-energy collision processes was first presented in 
\cite{Jentschura:2010ap,Jentschura:2011ih} by adapting the most general framework of \cite{Kotkin:1992bj}.
It was further developed in \cite{Ivanov:2011kk,Ivanov:2011bv,Karlovets:2012eu,Karlovets:2015nva,Karlovets:2016jrd,Karlovets:2018iww,Silenko:2019okz,Karlovets:2020odl},
see also the recent reviews on twisted electrons \cite{Bliokh:2017uvr,Lloyd:2017} and photons \cite{Paggett:2017,Knyazev:2019}.
	In this section, we will recapitulate this formalism, first for scalar twisted states and then for photons and electrons.
The calculations become most transparent for the so-called Bessel twisted states described below.
This is not the only option available. Collisions involving Laguerre-Gaussian twisted states
\cite{Karlovets:2018iww,Karlovets:2020odl} and other wave packets with non-trivial phase structure \cite{Karlovets:2016jrd}
also demonstrated remarkable features not present in (approximate) plane-wave collisions.
However for the purposes of the present paper, we find it sufficient to stay with (Gaussian-smeared) Bessel states.

A Bessel twisted state is a solution of the free wave equation 
with a definite energy $E$, longitudinal momentum $k_z$, 
modulus of the transverse momentum $|\bk_\perp|=\varkappa$ 
and a definite $z$-projection of the total angular momentum $m$, which must be integer.
Since, for the scalar field, the total angular momentum coincides with the orbital angular momentum (OAM),
the same parameter $m$ also quantifies the $z$-projection of the OAM.
Written in cylindric coordinates $\rho, \varphi_r, z$, this solution
$|E,\varkappa,m\rangle$ has the form
\be
|E,\varkappa, m\rangle = e^{-i E t + i k_z z} \cdot
\psi_{\varkappa m}(\br_\perp)\,, \quad \psi_{\varkappa m}(\br_\perp) = e^{i m \varphi_r} \sqrt{\varkappa \over 2\pi}J_{m}(\varkappa \rho)\,,
\label{twisted-coordinate}
\ee
where $J_m(x)$ is the Bessel function. 
This function is normalized according to
\begin{equation}
\int d^2\br_\perp \psi^*_{\varkappa' m'}(\br_\perp) \psi_{\varkappa m}(\br_\perp) 
= \delta(\varkappa-\varkappa')\delta_{m, m'}\,.\label{normalization}
\end{equation}
The azimuthal angle dependence $\propto e^{im\varphi_r}$ is the hallmark feature of the phase vortex.
A twisted state can be represented as a superposition of plane waves:
\be
|E,\varkappa,m\rangle = e^{-i E t + i k_z z} \int {d^2 \bk_\perp \over(2\pi)^2}a_{\varkappa m}(\bk_\perp) e^{i\bk_\perp \br_\perp}\,,
\label{twisted-def}
\ee
where
\be
a_{\varkappa m}(\bk_\perp)= (-i)^m e^{im\varphi_k}\sqrt{\frac{2\pi}{\varkappa}} \delta(|\bk_\perp|-\varkappa)\,.\label{a}
\ee
is the corresponding Fourier amplitude. This expansion can be inverted,
which means that twisted states form a complete basis for (transverse) wave functions \cite{Jentschura:2010ap,Jentschura:2011ih}.

When passing from plane waves to twisted states, one should also take care of the change
of the normalization factors. The accurate treatment of these factors can be found in  
\cite{Karlovets:2012eu,Serbo:2015kia}. 
Here, although the appropriate normalization coefficients are implicitly assumed,
and we do not write them for the following reason.
The absolute value of the twisted scattering cross section depends not only on the dynamics 
of the fundamental interactions but also on the details of how the initial
twisted state is prepared. These details depend on the eventual experimental realization
of the twisted states. Therefore, in contrast to the usual plane-wave setting,
the absolute value of the cross section cannot be unambiguously predicted.

	If one looks into the integrated cross section, its departure from the plane wave cross section 
is typically small and often negligible, see the very recent study \cite{Karlovets:2020odl}.
However the most dramatic novelties of the two-twisted-particle collision arise not in the absolute value of the cross section 
but in differential distributions absent in the plane wave case.
Since the absolute value of the cross section is not the figure of merit for the present study, we will often skip the normalization factors
and plot cross sections in arbitrary units.

We remark in passing that sometimes a different normalization of $a_{\varkappa m}(\bk_\perp)$ is adopted, 
namely, with the coefficient $2\pi/\varkappa$ instead of $\sqrt{2\pi/\varkappa}$.
This is the consequence of a different normalization condition for the coordinate wave function: with or without the prefactor
$2\pi/\varkappa$ in Eq.~\eqref{normalization}. This difference does not change the observables;
one just needs to keep track of the exact normalization choice when calculating the event rate and the flux.

If the above Bessel state describes a particle with mass $\mu$, the energy and momentum are related as
$E^2 = \mu^2 + \varkappa^2 + k_z^2$.
Notice that the {\em average} momentum of this state $\langle\bk\rangle = (0, 0, k_z)$
does not satisfy the dispersion relation:
$E^2 = \mu^2 + \varkappa^2 + \langle\bk\rangle^2 \not = \mu^2 + \langle\bk\rangle^2$.
Whether to interpret the quantity $\mu^2 + \varkappa^2$ as a new ``effective mass'' squared
is just a matter of terminological convenience.

Just like a plane wave, a pure Bessel state $|E,\varkappa,m\rangle$ with fixed $\varkappa$ 
is non-normalizable in the transverse plane. Although the resulting singularities can be dealt with
\cite{Jentschura:2010ap,Jentschura:2011ih,Ivanov:2011kk,Karlovets:2012eu},
it is more appropriate to use realistic, transversely localized 
monochromatic beams\footnote{We stress that a monochromatic solution with a localized transverse wave function
cannot be localized in the $z$ direction; otherwise, monochromaticity is lost.
Therefore such solutions correspond to {\em beams} rather than {\em wave packets},
although we will occasionally use the latter term as well.}. Such a beam can be written as a superposition of Bessel states 
with equal energy and equal values of $m$ but with a distribution over $\varkappa$,
\be
|E,\bar\varkappa,\sigma,m\rangle = \int d\varkappa \, f(\varkappa) |E,\varkappa,m\rangle\,.\label{WP}
\ee
The weight function $f(\varkappa)$ should be peaked at $\bar\varkappa$ and have a width $\sigma$; 
apart from that, it is unconstrained and will depend on details of a future experimental realization scheme.
In our calculations below, we will use the Gaussian function corrected by a slow-varying prefactor:
\begin{eqnarray}
f(\varkappa)=n \sqrt{\varkappa}\exp\left[
-\frac{(\varkappa-\bar\varkappa)^2}{2\sigma^2}\right]\,,\label{gaussian}
\end{eqnarray}
with the normalization condition $\int_0^{E} d\varkappa |f(\varkappa)|^2=1$.

\subsection{Description of twisted photons}

When describing  twisted photons, we adapt the formalism of 
\cite{Jentschura:2010ap,Jentschura:2011ih,Knyazev:2019}.
For definiteness, we will work in the Coulomb gauge, where all
polarization vectors only have the spatial components.
A monochromatic plane-wave electromagnetic field with helicity 
$\lambda = \pm 1$ is described by
\be
\bA_{\bk \lambda}(\br) = \bbe_{\bk \lambda}\, e^{i\bk \br}\,.\label{PW1}
\ee
The polarization vector is orthogonal to the wave vector: $\bbe_{\bk\lambda} \bk = 0$.
Quantization of this field produces plane wave photons with momentum $\bk$.

As for the scalar case, we fix a reference frame, select an axis $z$, and 
construct a Bessel twisted photon 
as a superposition of plane waves with fixed longitudinal momentum 
$k_z = |\bk|\cos\theta$, fixed modulus of the transverse momentum $\varkappa = |\bk_\perp| = k\sin\theta$,
but arriving from different azimuthal angles $\varphi_k$.
Such a twisted photon with a definite $z$-projection of the {\em total} angular momentum $m$
and definite helicity $\lambda = \pm 1$ can be written as
\be
\bA_{\varkappa m \lambda}(\br) = e^{i k_z z} \int a_{\varkappa m}(\bk_\perp)\, \bbe_{\bk \lambda}\, e^{i\bk_\perp \br_\perp} {d^2\bk_\perp \over (2\pi)^2}\,,
\label{tw1-ph}
\ee
where the Fourier amplitude $a_{\varkappa m}(\bk_\perp)$ is given by the same Eq.~\eqref{a}.
The usual dispersion relation holds for every plane wave component: $k_z^2 + \varkappa^2 = E^2$.

In contrast to the scalar case, each plane wave component of a twisted photon
contains its polarization vector $\bbe_{\bk \lambda}$, which is orthogonal 
to the momentum of that particular plane wave component: $\bbe_{\bk\lambda} \bk = 0$.
As a result, the polarization vector cannot be taken out of the integral. 
Back in the coordinate space, the polarization state of a twisted photon 
is described by a polarization {\em field} rather than a polarization vector.

To describe the polarization vector of a photon with an arbitrary momentum, 
let us define the eigenvectors ${\bm \chi}_\sigma$, $\sigma= \pm 1, 0$,
of the helicity operator $\hat{s}_z$ defined with respect to the axis $z$: $\hat{s}_z {\bm \chi}_\sigma = \sigma {\bm \chi}_\sigma$.
Their explicit form is 
\be
{\bm \chi}_{0}=
\left(
\begin{tabular}{c}
	0 \\
	0 \\
	1 \\
\end{tabular}
\right),\ 
{\bm \chi}_{\pm 1}= \fr{\mp 1}{\sqrt{2}}
\left(
\begin{tabular}{c}
	1 \\
	$\pm i$ \\
	0 \\
\end{tabular}
\right)\,, \quad {\bm \chi}^*_\sigma {\bm \chi}_{\sigma^\prime} = \delta_{\sigma\sigma^\prime}\,.
\label{chi}
\ee
The polarization vector can be expanded in the basis of ${\bm \chi}_\sigma$:
\be
\bbe_{\bk \lambda}=
\sum_{\sigma=0,\pm 1} e^{-i\sigma \varphi_k}\,
d^{1}_{\sigma \lambda}(\theta)  \,\bm \chi_{\sigma}\,.
\label{bbe-chi}
\ee
The explicit expressions for Wigner's $d$-functions~\cite{LL3} are:
\be
d^{1}_{\sigma\lambda} = \mmmatrix{\cos^2\fr{\theta}{2}}{-\fr{1}{\sqrt{2}}\sin\theta}{\sin^2\fr{\theta}{2}}%
{\fr{1}{\sqrt{2}}\sin\theta}{\cos\theta}{-\fr{1}{\sqrt{2}}\sin\theta}%
{\sin^2\fr{\theta}{2}}{\fr{1}{\sqrt{2}}\sin\theta}{\cos^2\fr{\theta}{2}} .
\ee
The first, second, and third rows and columns of this matrix correspond to the indices $+1,\, 0,\, -1$.
Performing the summation in Eq.~\eqref{bbe-chi}, one gets explicit expressions for the polarization vectors:
\be
\bbe_{\bk \lambda}= \fr{\lambda }{\sqrt{2}}\triplet{-\cos\theta\cos\varphi_k + i \lambda \sin\varphi_k}%
{-\cos\theta\sin\varphi_k - i \lambda \cos\varphi_k}{\sin\theta}\,,\quad \lambda= \pm 1\,.\label{bbe-explicit}
\ee

Notice that the Fourier amplitude $a_{\varkappa m}(\bk_\perp)$ is an eigenstate 
not only of $\hat{J}_z$, the operator of the $z$-component of the total angular momentum,
but also of $\hat{L}_z = -i \partial/\partial \varphi_k$, the $z$-projection of the OAM operator.
However, this property is not shared by $\bbe_{\bk \lambda}$ given above:
it is an eigenstate of $\hat{J}_z$ with the zero eigenvalue but not of $\hat{L}_z$ or $\hat{s}_z$ separately.
This polarization vector, even for fixed $\lambda$, contains contributions with different $s_z$ and $\ell = L_z$,
which sum up to zero.
Thus, the twisted photon \eqref{tw1-ph}, strictly speaking, is not an eigenstate of the OAM 
because the spin and OAM projections are not conserved separately 
even for free electromagnetic fields. 

In most experimental situations, twisted photons are produced in the paraxial regime,
where $\theta \ll 1$. In this case, one can talk about approximately conserved $s_z = \lambda$ 
and $\ell = m - \lambda$.
Indeed, when $\theta\to 0$, the polarization vector becomes
\be
\bbe_{\bk \lambda} \to e^{-i\lambda \varphi_k} \,\bm \chi_{\lambda}\,,
\ee
which now has definite $s_z = - \ell = \lambda$.
Beyond the paraxial approximation, the spin-orbital interaction, which exists for free electromagnetic waves,
comes into play and gives rise to a variety of remarkable optical phenomena \cite{Bliokh:2015yhi}.
In particular, it leads to spatially varying polarization states of light described by polarization field.
In a tightly focused light beam, the polarization field evolves downstream
and may significantly differ at the aperture and in the focal plane.

Finally, when describing a counter-propagating twisted photon defined in the same reference frame
with respect to the same axis $z$, one can use the above expressions assuming that $k_z < 0$
and replacing $m \to - m$ in the Fourier amplitude \eqref{a}.
The expression for the polarization vector \eqref{bbe-explicit} stays unchanged,
but $\cos\theta < 0$. The paraxial limit is now given by $\theta \to \pi$, in which case 
$\bbe_{\bk \lambda} \to e^{+i\lambda \varphi_k} \,\bm \chi_{-\lambda}$.

\subsection{Description of twisted electrons and positrons}

Twisted states have been experimentally demonstrated not only for photons
but also for electrons \cite{Uchida:2010,Verbeeck:2010,McMorran:2011}. 
To describe them in a fully relativistic manner, 
we use the definitions of \cite{Serbo:2015kia,Bliokh:2017uvr};
other works, such as \cite{Bliokh:2011fi,Karlovets:2012eu}, use slightly different conventions.
The plane-wave electron with the four-momentum
$k^\mu = (E,\, \bk_\perp,\, k_z)$, corresponding to the propagation direction with angles
$\theta$ and $\varphi_k$, and with helicity $\zeta = \pm 1/2$ is described by
\be
\Psi_{k \zeta}(\br)= {1 \over \sqrt{2E}}\, u_{\zeta}(k)\, e^{i\bk \br}\,.
\label{PW}
\ee
The bispinor $u_{\zeta}(k)$ used here is 
\be
u_{\zeta}(k) = \doublet{\sqrt{E+m_e}\,w^{(\zeta)}}{2 \zeta \sqrt{E-m_e}\,w^{(\zeta)}}\,, \quad 
w^{(+1/2)} = \doublet{c\, e^{-i\varphi_k/2}}{s\, e^{i\varphi_k/2}}\,,
\quad w^{(-1/2)} = \doublet{-s\, e^{-i\varphi_k/2}}{c\,e^{i\varphi_k/2}}\,,\label{PWspinors}
\ee
where $c \equiv \cos(\theta/2)$, $s \equiv \sin(\theta/2)$.
The bispinors are normalized as 
$\bar u_{\zeta_1}(k) u_{\zeta_2}(k) = 2m_e\, \delta_{\zeta_1, \zeta_2}$.
The negative-frequency solutions of the Dirac equation are constructed as
\be
v_{\zeta}(k) = \doublet{-\sqrt{E-m_e}\,w^{(-\zeta)}}{2\zeta \sqrt{E+m_e}\,w^{(-\zeta)}}\,,
\ee
with the same spinors $w$ as in \eqref{PWspinors}.
We use this basis of plane-wave solutions of the Dirac equation to construct the Bessel vortex state of the electron:
\be
\label{bessel}
\Psi_{\varkappa m k_z \zeta}(\br) = e^{i k_z z} 
\int a_{\varkappa m}(\bk_\perp)\, {u_{\zeta}(k) \over \sqrt{2E}} \, e^{i\bk_\perp \br_\perp} {d^2\bk_\perp \over (2\pi)^2}\,,
\ee
with the same Fourier amplitude $a_{\varkappa m}(\bk_\perp)$ as in Eq.~\eqref{a}.
Notice that the total angular momentum projection $m$ is now half-integer.
The similar expression holds for the negative-frequency solutions.

Just as in the case of twisted photons, the spin and orbital angular momentum projections 
are not separately conserved due to the intrinsic spin-orbital interaction 
of the twisted electron, \cite{Bliokh:2011fi,Karlovets:2018iww}.
In the paraxial approximation, when the spin-orbital interaction is suppressed,
one can nevertheless talk about two approximately conserved quantum numbers: 
the $z$ projection of the spin operator with $s_z = \zeta$
and the $z$-projection of the OAM with $\ell = m - \zeta$.
One could also define Bessel electron states in which the spinor $u_{k\zeta}$ contains 
an extra factor $\exp(i\zeta\varphi_k)$, while the Fourier amplitude (\ref{a}) is constructed with integer $\ell$
instead of half-integer $m$ \cite{Karlovets:2012eu}. This is also a valid Bessel electron solution; 
its total angular momentum depends on helicity, $m = \ell + \zeta$,
while the parameter $\ell$ characterizes the orbital angular momentum independent of helicity. 
These two conventions correspond to two definitions of how an unpolarized electron is defined, see below.

\subsection{Unpolarized twisted photons or electrons}\label{subsection-unpolarized}

Let us discuss the subtle notion of {\em unpolarized} twisted photons. 
For concreteness, we talk about photons, but 
the entire discussion is applicable to electrons and other particles with spin. 

Due to the presence of spin-orbital interaction of light in free space,
the notion of unpolarized twisted light is not unambiguously defined.
For a plane-wave photon, with its polarization vector independent of spatial coordinates,
one can think of unpolarized light as an equal mixture of photons
in two orthogonal polarization states, for examples, with $\lambda= +1$
and $\lambda=-1$. For twisted light, an ambiguity arises:
when considering photons with $\lambda = \pm 1$, 
should we keep the total angular momentum $m$ fixed?
Or should we fix $m-\lambda$, which would correspond 
in the paraxial limit to the same spatial distribution
of the two polarization states?

There is no unique answer to this question;
it will depend on the photon preparation details in every experimental scheme.
If an experimental device manages to select photons 
with a single $m$ irrespective of the photon helicity,
then one needs to calculate the process of interest 
(in our case, the cross section) with $|m, \lambda = +1\rangle$
and $|m, \lambda = -1\rangle$ and perform the averaging.
If one creates twisted photons by letting them pass through
a fixed aperture plate which would impose a given OAM $\ell$ in the scalar case,
then, immediately behind the aperture, 
one can reliably describe the unpolarized twisted light
as consisting of photons with $|m_+ = \ell + 1, \lambda = +1\rangle$
and $|m_- = \ell - 1, \lambda = -1\rangle$.
However, this description evolves downstream and may become very different in the focal spot 
due to the same spin-orbital interaction of light.
In particular, it leads to spin-to-orbital conversion which was experimentally verified in \cite{beads2007}
by shining left or right circularly polarized light to the same numerical aperture
and observing that an ensemble of microscopic target particles in the focal plane 
rotated as a whole in different manner in these two cases due to the different amount
of OAM in the focal plane.

It means that when calculating processes with unpolarized twisted photons in realistic settings, 
one must specify according to which definition the twisted light is unpolarized.
It is well possible that the realistic situation will correspond to an intermediate definition 
between the two options just described.
In the next section, when describing twisted photon collisions,
we will discuss how the results differ with the two definitions of unpolarized twisted photons:
fixed-$m$ and fixed-$\ell$ options. 
We will show that with both definitions, the key spin- or parity-sensitive 
observables do not vanish, although their magnitude will be different.
Their value in a realistic experimental situation will likely lie in between.


\section{Resonance production in twisted photon collisions}\label{section-photons}

\subsection{General features of twisted particle annihilation}

Let us begin by briefly recapitulating the main features of two twisted particles collisions, see more details in \cite{Ivanov:2019pdt}.

If we were to describe a $2\to 1$ annihilation process for the plane wave case, 
we would need to write the $S$-matrix amplitude as
\be
S_{PW} = i(2\pi)^4\delta(E_1+E_2-E_K) \delta^{(3)}(\bk_1 + \bk_2 - \bK) {{\cal M}(k_1,k_2;K) \over \sqrt{8 E_1 E_2 E_K}}\,.
\label{SPW}
\ee
Here the energies and momenta of the initial particles are $E_i$ and $\bk_i$,
for the final particle $E_K$ and $\bK$. ${\cal M}(k_1,k_2;K)$ is the invariant amplitude 
calculated according to the standard Feynman rules.
Squaring this amplitude and appropriately regularizing the squares of delta-functions, as described, for instance, in \cite{LL4},
we would get the cross section 
\bea
d\sigma &=& {\pi \delta(E_1+E_2-E_K) \over 4 E_1 E_2 E_K v} |{\cal M}|^2  \, 
\delta^{(3)}(\bk_1 + \bk_2 - \bK) \, d^3 K\,,\nonumber\\[2mm]
\sigma &=& {\pi \delta(E_1+E_2-E_K) \over 4 E_1 E_2 E_K v} |{\cal M}|^2\,.\label{sigma-PW-0}
\eea
Notice the well known features of this cross section:
the final momentum is fixed at $\bK = \bk_1 + \bk_2$, and the dependence
on the total energy of the colliding particles is proportional to $\delta(E_1+E_2-E_K)$.
The production process occurs only when the initial particles are directly ``on resonance''.

Let us now consider collision of two Bessel states $|E_1,\varkappa_1,m_1\rangle$
and $|E_2,\varkappa_2,m_2\rangle$ which are defined with respect to the same axis $z$.
The final particle with mass $M$ is still described in the basis of plane waves
with the momentum $\bK$ and energy $E_K$.
The $S$-matrix element of this process is 
\be
S = \int {d^2 \bk_{1\perp} \over (2\pi)^2} {d^2 \bk_{2\perp} \over (2\pi)^2} 
a_{\varkappa_1 m_1}(\bk_{1\perp}) a_{\varkappa_2, -m_2}(\bk_{2\perp}) S_{PW} = 
i (2\pi)^4 \fr{\delta(\Sigma E) \delta(\Sigma k_z)}{\sqrt{8 E_1 E_2 E_K}} 
{(-i)^{m_1-m_2} \over (2\pi)^3\sqrt{\varkappa_1\varkappa_2}} \cdot {\cal J}\,,\label{S-tw}
\ee
where $\delta(\Sigma E) \equiv \delta(E_1+E_2-E_K)$, $\delta(\Sigma k_z) \equiv \delta(k_{1z}+k_{2z}-K_z)$.
The twisted amplitude ${\cal J}$ is defined as
\bea
{\cal J} &=& \int d^2 \bk_{1\perp} d^2 \bk_{2\perp} \, e^{im_1\varphi_1 - im_2\varphi_2}\,
\delta(|\bk_{1\perp}|-\varkappa_1) \delta(|\bk_{2\perp}|-\varkappa_2)
\delta^{(2)}(\bk_{1\perp}+\bk_{2\perp} - \bK_\perp)\cdot {\cal M}\nonumber\\
&=&\varkappa_1\varkappa_2 \int d\varphi_1 d\varphi_2\, e^{im_1\varphi_1 - im_2\varphi_2}\,
\delta^{(2)}(\bk_{1\perp}+\bk_{2\perp} - \bK_\perp)\cdot {\cal M}\,.\label{J}
\eea
\begin{figure}[ht]
\centering
\includegraphics[width=0.75\textwidth]{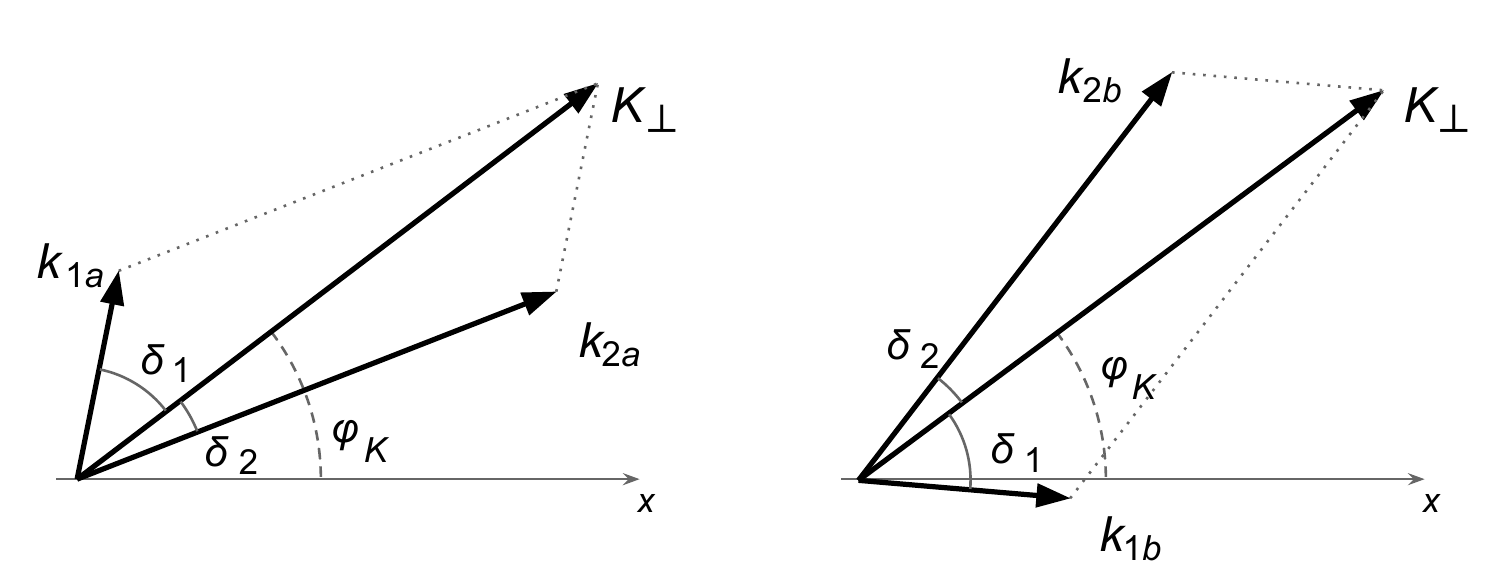}
{\caption{\label{fig-2configurations} The two kinematical configurations in the transverse plane that satisfy
momentum conservation laws in the scattering of two Bessel electron states.}}
\end{figure}
Remarkably, the twisted amplitude ${\cal J}$ can be calculated exactly \cite{Ivanov:2011kk}. 
It is non-zero only if $\varkappa_i$ and $K \equiv |\bK_\perp|$ satisfy the triangle inequalities
\be
|\varkappa_1 - \varkappa_2| \le K \le \varkappa_1 + \varkappa_2\,.\label{ring}
\ee
They form a triangle with the area
\be
\Delta 
= {1 \over 4} \sqrt{2 K^2\varkappa_1^2 + 2 K^2\varkappa_2^2 + 2\varkappa_1^2\varkappa_2^2 
- K^4 - \varkappa_1^4 - \varkappa_2^4}\,.\label{area}
 \ee
Out of many plane wave components ``stored'' in the initial twisted particles, 
the integral \eqref{J} receives contributions from exactly two plane wave combinations shown in Fig.~\ref{fig-2configurations}
with the following azimuthal angles:
\bea
\mbox{configuration a:} &&\varphi_1 = \varphi_{K} + \delta_1\,,\quad \varphi_2 = \varphi_{K} - \delta_2\,,\nonumber\\
\mbox{configuration b:} &&\varphi_1 = \varphi_{K} - \delta_1\,,\quad \varphi_2 = \varphi_{K} + \delta_2\,.\label{phi12}
\eea
Notice that
\be
\delta_1 = \arccos\left({\varkappa_1^2 + K^{2} - \varkappa_2^2 \over 2\varkappa_1 K}\right)\,,
\quad
\delta_2 = \arccos\left({\varkappa_2^2 + K^{2} - \varkappa_1^2 \over 2\varkappa_2 K}\right)
\label{delta_i}
\ee
are the inner angles of the triangle with the sides $\varkappa_1$, $\varkappa_2$, $K$;
they are not azimuthal variables. 

The result for the twisted amplitude ${\cal J}$ can then be compactly written as
\be
{\cal J} = e^{i(m_1 - m_2)\varphi_{K}}{\varkappa_1 \varkappa_2 \over 2\Delta}
\left[{\cal M}_{a}\, e^{i (m_1 \delta_1 + m_2 \delta_2)} + {\cal M}_{b}\, e^{-i (m_1 \delta_1 + m_2 \delta_2)}\right]\,.\label{J2}
\ee
Notice that the plane-wave amplitudes ${\cal M}_{a}$ and ${\cal M}_{b}$ are calculated for the two distinct initial momentum 
configurations shown in Fig.~\ref{fig-2configurations} but for the same final momentum $\bK$.
They exhibit two distinct paths in momentum space to arrive at the same final-state from the initial twisted states.
In a sense, scattering of twisted Bessel states represents the momentum-space analog of the Young double-slit
experiment \cite{Ivanov:2016jzt}.

Squaring \eqref{S-tw} and performing appropriate regularization,
we obtain the (generalized) cross section in the form
\be
d\sigma \propto |{\cal J}|^2 \delta(E_1+E_2-E_K) \, d^2\bK_\perp\,.\label{dsigma-tw}
\ee
We deliberately omitted the prefactor to stress, as we discussed in the previous section,
that the absolute value of the cross section cannot be predicted unambiguously
as it depends on the details of initial state preparation and, therefore, it is not the figure of merit.
Instead, it is the non-trivial distribution over $\bK_\perp$, 
which was absent in the plane wave case \eqref{sigma-PW-0},
that we pay attention to.

For fixed initial values of $E_i$, $\varkappa_i$, and $M$, the energy-momentum conservation 
fixes $K_z = k_{1z}+k_{2z}$ and, therefore, the modulus of the transverse momentum $K = \sqrt{E_K^2 - M^2 - K_z^2}$. 
Thus, the polar angle of the produced resonance is fixed \cite{Ivanov:2019pdt}: 
\begin{equation}
\cos\theta_K = \fr{K_z}{\sqrt{(E_1+E_2)^2-M^2}}\,,\label{theta-K}
\end{equation}
but the cross section exhibits a uniform distribution in the azimuthal angle.

The expression for the cross section \eqref{dsigma-tw} and the exact evaluation of ${\cal J}$ in \eqref{J2}
were obtained for the pure Bessel states, which are not normalizable and lead to singularities in the cross sections.
These singularities are removed for realistic twisted wave packets with a finite transverse extent, 
for which we use the monochromatic Gaussian-smeared wave packet given in Eqs.~\eqref{WP}, \eqref{gaussian}.
This smearing with the functions $f_1(\varkappa_1)$ and $f_2(\varkappa_2)$ must be applied at the level of $S$-matrix amplitude \eqref{S-tw}.
Therefore, instead of pure Bessel twisted amplitude ${\cal J}$, we evaluate its smeared counterpart:
\be
\lr{{\cal J}} = \int d\varkappa_1 d\varkappa_2 f_1(\varkappa_1) f_2(\varkappa_2) \delta(k_{1z} + k_{2z} - K_z) 
{{\cal J} \over \sqrt{\varkappa_1\varkappa_2}}\,.\label{J-smeared}
\ee
Notice that this integration now affects the longitudinal momenta, since, for monoenergetic states,
variation of $\varkappa$ induces variation of $k_z$.
Therefore, the final particle now displays a 2D momentum space distribution,
which can be written as 
\be
d\sigma \propto E_K^2 \beta_K\, |\lr{{\cal J}}|^2\, d\Omega_K = {E_K \over K_z}\, |\lr{{\cal J}}|^2\, d^2 \bK_\perp\,.\label{dsigma-tw-3}
\ee
Further insights into this distribution can be found in \cite{Ivanov:2019pdt}.

\subsection{Scalar resonance production in twisted $\gamma\gamma$ collisions}

\subsubsection{Exact expressions}

Production of a spin-0 resonance in twisted $\gamma\gamma$ collision can be described 
with the same formalism as in the scalar case, corrected for the presence of polarization vectors \cite{Ivanov:2019lgh}.
One encounters the same twisted amplitude ${\cal J}$ as
in \eqref{J} and \eqref{J2}, where $m_i$ now refer to the total angular momentum of each photon,
while the invariant amplitude ${\cal M}$ depends now on the photon helicities.
To calculate it, suppose $S$ is a real scalar field which can be produced in $\gamma\gamma$ collision
through the following effective interaction Lagrangian:
\be
{\cal L}_S = {g \over 4} F^{\mu\nu} F_{\mu\nu} S\,.
\ee
It generates the following helicity amplitude
\be
{\cal M}_S = g\left[ (k_1k_2)(e_1e_2) - (k_1e_2)(k_2e_1)\right]\,,\label{Ms} 
\ee
where all products are understood as 4-vector products.
For plane-wave collisions, one usually chooses the center of motion frame,
in which the polarization vectors, written in the Coulomb gauge, are orthogonal to the momenta
of {\em both} photons, which allows one to drop the second term in \eqref{Ms}.
In our case, this orthogonality does not hold in the Coulomb gauge, and both terms must be evaluated
in the reference frame we work in.

Using the explicit expressions for the polarization vectors and momenta, one can evaluate
the products entering this expression:
\bea
(k_1k_2) &=& {1\over 2}M^2 = E_1E_2[1-c_1c_2-s_1s_2\cos(\varphi_1-\varphi_2)]\,,\nonumber\\
(e_1 e_2) &=& e^{i(\varphi_1-\varphi_2)}{1-\lambda_1 c_1 \over 2}{1+\lambda_2 c_2 \over 2} 
+ e^{-i(\varphi_1-\varphi_2)}{1+\lambda_1 c_1\over 2}{1-\lambda_2 c_2\over 2} - {\lambda_1 \lambda_2 s_1 s_2 \over 2}\,,
\nonumber\\
(e_1 k_2) &=& {E_2 \over \sqrt{2}} \left[e^{i(\varphi_1-\varphi_2)} {\lambda_1 c_1 -1 \over 2} s_2  
+ e^{-i(\varphi_1-\varphi_2)}{1+\lambda_1 c_1\over 2}s_2 - \lambda_1 s_1 c_2\right]\,,
\eea
and similarly for $(e_2k_1)$. 
We adopted here the shorthand notation $c_i \equiv \cos\theta_i$, $s_i \equiv \sin\theta_i$.
Notice that the plane wave amplitude depends on the azimuthal angles 
of the two photons only through their difference: 
${\cal M}_S(\varphi_1, \varphi_2) = {\cal M}_S(\varphi_1- \varphi_2)$.
Substituting these products into \eqref{Ms} and simplifying the expressions, 
we get a non-zero amplitude only for equal helicities $\lambda_1=\lambda_2=\lambda$:
\bea
{\cal M}_S &=& 2 g E_1E_2 \delta_{\lambda_1, \lambda_2}\left[
e^{i(\varphi_1-\varphi_2)}{1-\lambda c_1 \over 2}{1+\lambda c_2 \over 2} 
+ e^{-i(\varphi_1-\varphi_2)}{1+\lambda c_1\over 2}{1-\lambda c_2\over 2} - {s_1 s_2 \over 2}\right]\nonumber\\[1mm]
&=& 2 g E_1E_2 \delta_{\lambda_1, \lambda_2} (e_1^{(\lambda)} e_2^{(\lambda)})\,.
\eea
Next, one calculates ${\cal J}$ via \eqref{J2}. The two interfering configurations 
differ only by their azimuthal angles: $\varphi_1-\varphi_2 = \delta_1 + \delta_2$ or $-(\delta_1 + \delta_2)$.
Thus, we get
\bea
{\cal J}_S &=& e^{i(m_1 - m_2)\varphi_{K}}{\varkappa_1 \varkappa_2 \over 2\Delta}\cdot g E_1E_2 \delta_{\lambda_1, \lambda_2} \times \nonumber\\
&&\times
\Bigl\{(1-\lambda c_1)(1+\lambda c_2)\cos\left[m_1\delta_1 + m_2\delta_2 + \delta_1 +\delta_2\right]
\nonumber\\[2mm]
&&\ \ + (1+\lambda c_1)(1-\lambda c_2)\cos\left[m_1\delta_1 + m_2\delta_2 - (\delta_1 +\delta_2)\right]
\nonumber\\[1mm]
&&\ \  - 2 s_1 s_2 \cos\left[m_1\delta_1 + m_2\delta_2\right]\Bigr\}\,.\label{JS1}
\eea
In the paraxial limit, when $\theta_1 \to 0$ meaning $c_1 \to 1$ and $\theta_2 \to \pi$ meaning $c_2 \to -1$, 
the first term dominates for $\lambda = -1$, while the second term dominates for $\lambda = +1$. 
The azimuthal angle dependence $\exp[i(m_1 - m_2)\varphi_{K}]$ indicates that the total angular momentum
of the initial two-photon system is converted in the OAM of the single final scalar particle, should we want 
to describe the latter in the basis of twisted states as well \cite{Ivanov:2019lgh}.

\subsubsection{Helicity dependence: fixed-$m$ case}

A remarkable feature of ${\cal J}_S$ seen in Eq.~\eqref{JS1} 
is its non-trivial dependence on helicity $\lambda = \lambda_1 = \lambda_2$.
Following the discussion in Section~\ref{subsection-unpolarized},
we now specify that, when considering unpolarized cross section, 
we fix $m_1$, $m_2$ and vary $\lambda = \pm 1$.
Then, the expression ${\cal J}_S$ can be written as
\be
{\cal J}_S = e^{i(m_1 - m_2)\varphi_{K}} \delta_{\lambda_1, \lambda_2} 
{gE_1E_2 \varkappa_1 \varkappa_2 \over \Delta}\cdot \left({\cal J}_1 + \lambda {\cal J}_2\right)\,,
\label{JS2}
\ee
where the real quantities ${\cal J}_1$ and ${\cal J}_2$ are
\bea
{\cal J}_1 &=& \cos(m_1\delta_1 + m_2\delta_2) \left[\cos(\delta_1 +\delta_2)(1-c_1c_2) - s_1 s_2\right]\,,
\nonumber\\
{\cal J}_2 &=& \sin(m_1\delta_1 + m_2\delta_2) \sin(\delta_1 +\delta_2) (c_1 - c_2)\,.\label{JS3}
\eea
This dependence survives in the cross section, which can be generically represented as
\be
\sigma_\lambda = \sigma_0 + \lambda \sigma_a \equiv \sigma_0(1+\lambda A)\,,\label{dsigma-ph-3}
\ee
with $\sigma_0$ representing the unpolarized cross section and $\sigma_a$ denoting the spin asymmetry.
The quantity $A \equiv \sigma_a/\sigma_0$ can be called the asymmetry contrast.
For the pure Bessel beams, this asymmetry contrast is given by
\be
A = {2{\cal J}_1{\cal J}_2 \over {\cal J}_1^2 + {\cal J}_2^2}\,,\label{asymmetry-contrast}
\ee 
which can vary between $-1$ and $+1$ depending on the exact position 
on the interference fringe.

\begin{figure}[!htb]
	\centering
	\includegraphics[height=5cm]{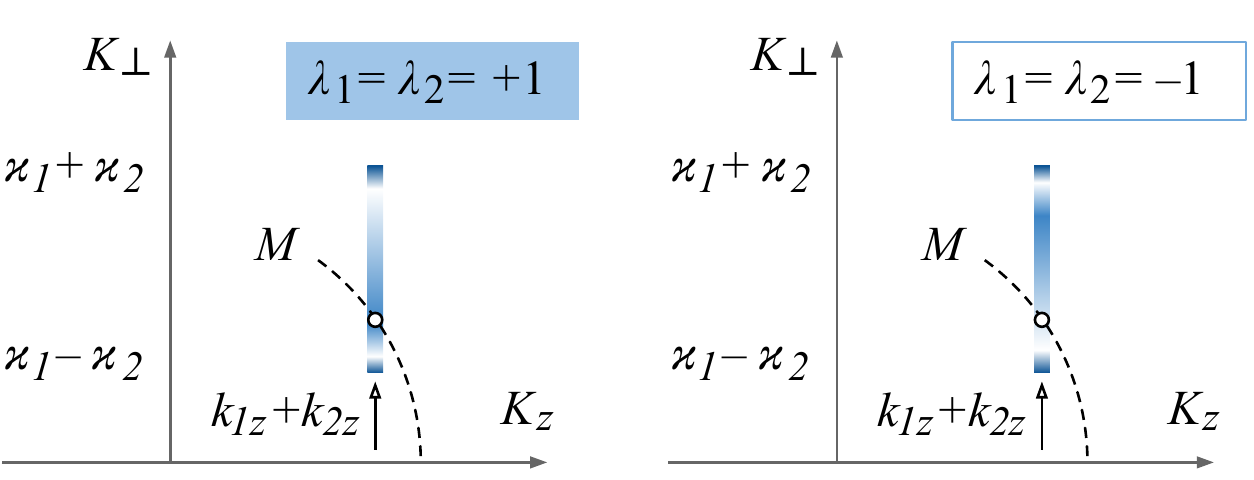}
	\caption{The absolute value of the transverse momentum $K$ and the longitudinal momentum $K_z$ of the produced scalar resonance of mass $M$ 
		in Bessel photon collision is uniquely reconstructed from the energy-momentum conservation for either helicity arrangement: 
		$\lambda_1 = \lambda_2 = +1$ (left plot) and $\lambda_1 = \lambda_2 = -1$ (right plot).
		However the magnitude of the cross section (shown as shades of blue) can be different due to different interference patterns, 
		leading to an energy-dependent polarization asymmetry.
	}
	\label{fig:2d-photons}
\end{figure}

This dependence may at first look surprising. Indeed, in the familiar plane wave collision,
the helicity combinations $\lambda_1 = \lambda_2 = +1$ and $\lambda_1 = \lambda_2 = -1$
lead to identical cross sections.
This is due to the fact the entire process is parity-invariant.
However in the presence case, we {\em explicitly break the left-right symmetry of the initial state} by selecting
twisted photons with definite values of $m_1$ and $m_2$.
A different pair of photons with angular momenta $-m_1$ and $-m_2$ 
would flip the sign of ${\cal J}_2$ in \eqref{JS3} and, consequently, 
in the asymmetry \eqref{asymmetry-contrast}.
In short, production of a scalar particle in twisted photon collision is invariant under the {\em simultaneous} sign flips 
$m_i \to -m_i$ and $\lambda_i \to - \lambda_i$,
but not under $\lambda_i \to - \lambda_i$ alone.

We stress that, for fixed $m$, the asymmetry does not vanish in the paraxial limit.
Indeed, setting $c_1 \to 1$, $c_2 \to -1$, $s_i \to 0$ in the above expressions, we obtain
\be
{\cal M}_S = 2 g E_1E_2 \delta_{\lambda_1, \lambda_2}  e^{-i\lambda(\varphi_1-\varphi_2)}\,,
\ee
and, as the result, we get 
\be
{\cal J}_S \propto \cos[(m_1-\lambda)\delta_1 + (m_2-\lambda)\delta_2] = \cos(\ell_1\delta_1 + \ell_2\delta_2)\,,\label{JS4}
\ee 
which coincides with the expression for twisted scalar annihilation \cite{Ivanov:2019pdt}.
Here, $\ell_i$ are the $z$ projections of the OAM of the two photons, 
which are approximately conserved in this limit.
The sizable spin asymmetry $\sigma_a$ originates from the fact that, in the paraxial approximation, 
$\lambda=+1$ involves the OAM state $\ell = m-1$, while $\lambda=-1$ involves 
the OAM state $\ell' = m+1 = \ell+2$. The two states have different spatial distributions.
As schematically illustrated by Fig.~\ref{fig:2d-photons}, 
the resonance production amplitude in collision of Bessel twisted photons
involves interference between two plane-wave amplitudes.
Since this interference depends on the OAM values,
one observes non-identical cross sections $\sigma_{\lambda = +1}$ and $\sigma_{\lambda = -1}$.

For Bessel photon collisions, if the initial kinematics is fixed, then the longitudinal momentum of the produced particle $K_z$
and its modulus of the transverse momentum $K$ are also fixed. This uniquely defines $\delta_1$ and $\delta_2$ and,
therefore, the exact position with respect to the interference fringes.
However if one performs the total energy scan, then $K$ and/or $K_z$ will vary,
and one can slide across interference fringes and observe rapidly changing asymmetry $A$.
In particular, one can choose a particular position on the fringe to enhance the asymmetry contrast $A$
as much as possible, that is, to achieve $\sigma_{\lambda = +1} \gg \sigma_{\lambda = -1}$
or vice versa. On the other hand, just as in the scalar case \cite{Ivanov:2019pdt}, 
one can anticipate that the Gaussian smearing of the pure Bessel states
will reduce fringe visibility and the asymmetry contrast.

\begin{figure}[h]
	\centerline{
		\includegraphics[width=0.45\textwidth]{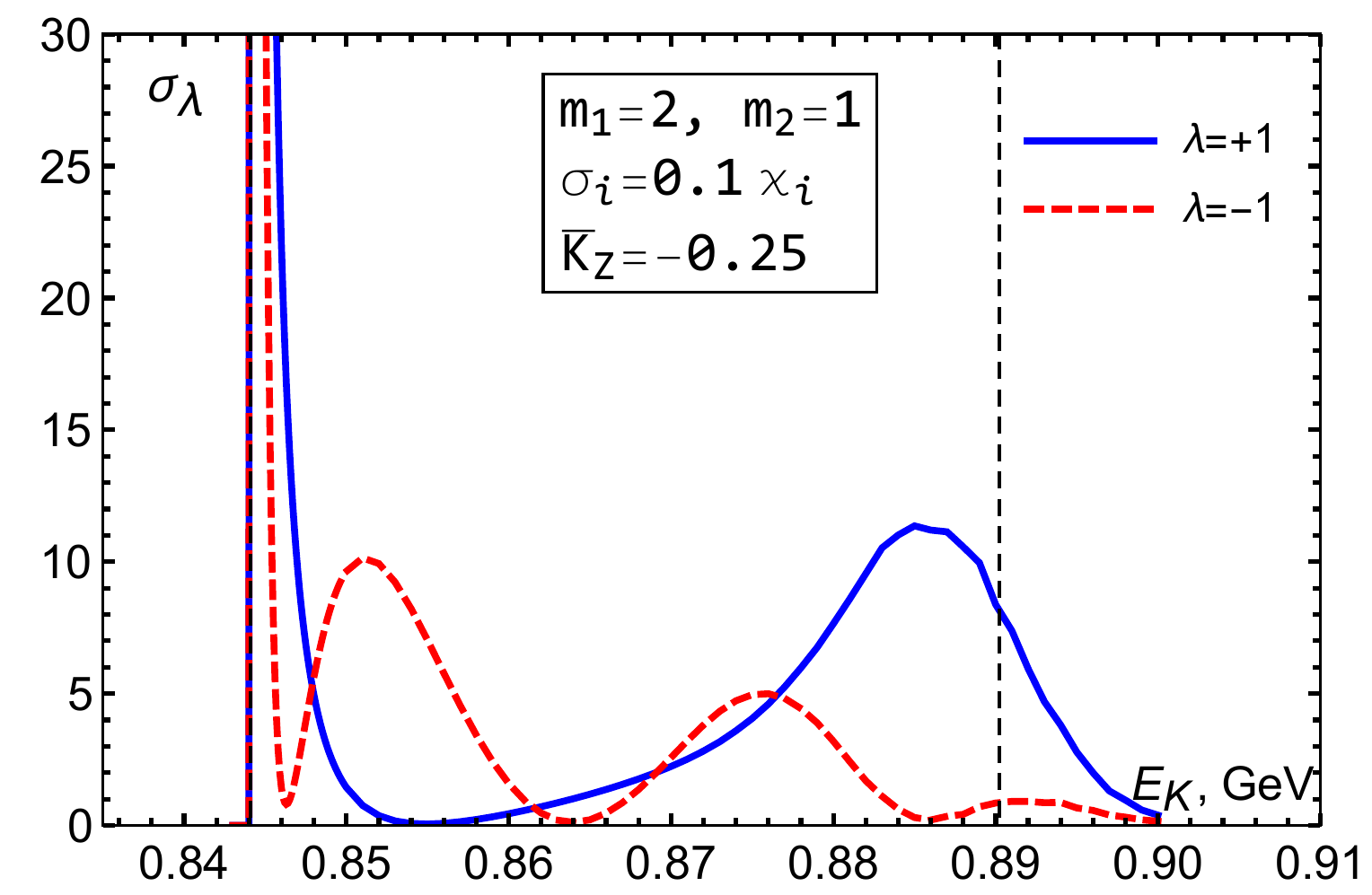}\hfill
		\includegraphics[width=0.45\textwidth]{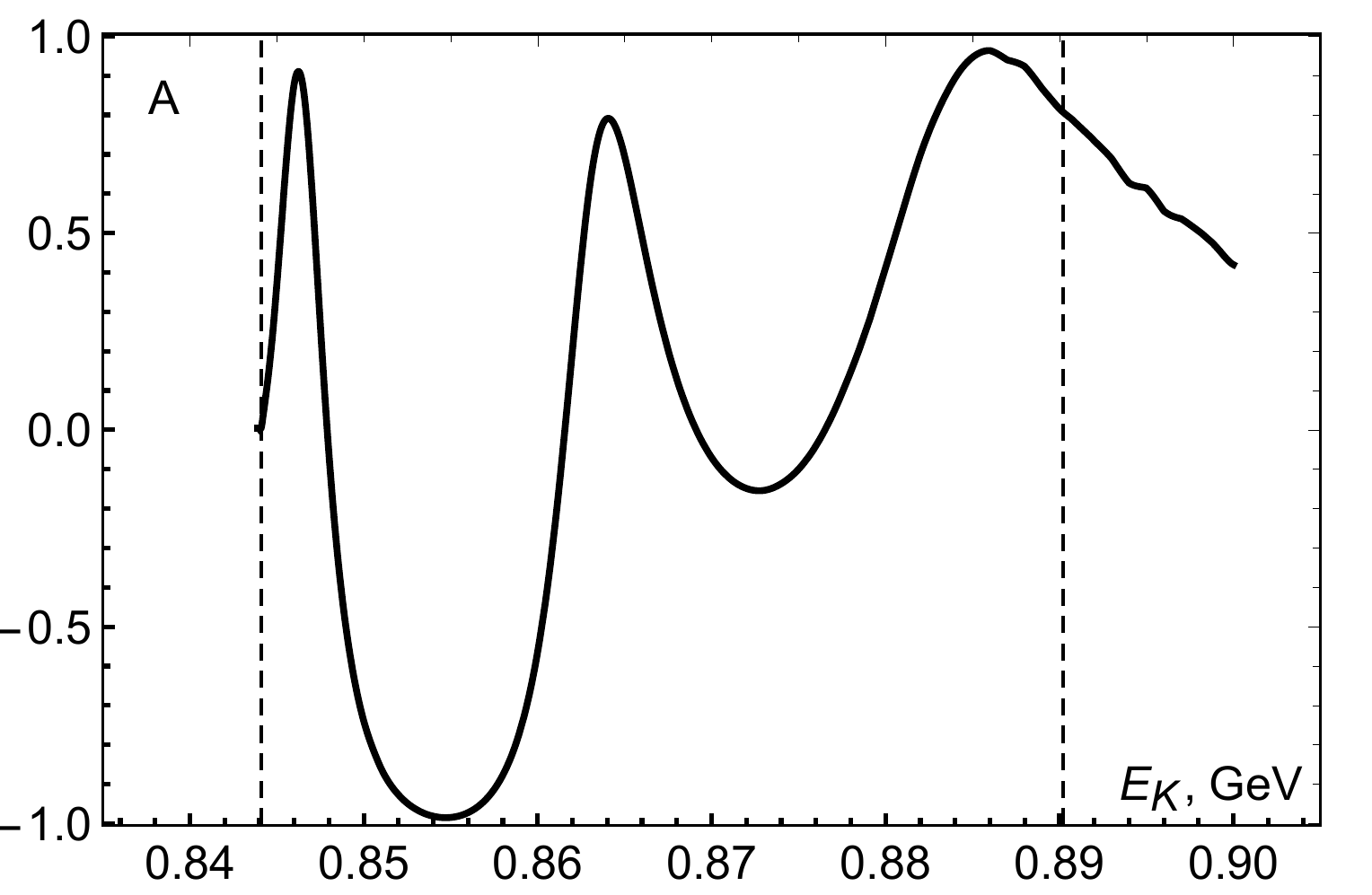}
	}
	\caption{
		\label{fig:gamgam-S}
		Left: Cross section of the process $\gamma\gamma\to S$ with twisted photons with parameters \eqref{gamgan-parameters} 
		for two helicity configurations:
		$\lambda_1 = \lambda_2 = +1$ (blue) and $-1$ (red). 
		Right: the asymmetry contrast $A$ defined in Eq.~\eqref{asymmetry-contrast}
	}
\end{figure}

	Fig.~\ref{fig:gamgam-S} demonstrates the typical values of the asymmetry one can obtain for realistic twisted photon beams.
	For illustration, we take a narrow resonance with mass $M= 0.8$ GeV and show the distribution of the cross sections $\sigma_\lambda$
	and the asymmetry contrast $A$ over the total energy $E_K$ for the Gaussian-smeared twisted states with kinematic parameters
	\begin{equation}
	(m_1, m_2) = (2, 1)\,, \quad \bar{\varkappa}_{1}=0.1~\GeV, \quad  \bar{\varkappa}_{2}=0.2~\GeV, 
	\quad \bar{K_z} = -0.25~\GeV\,, \label{gamgan-parameters} 
	\end{equation}
	and with $\sigma_i = \bar\varkappa_i/10$.
	Here and below we use the notation
	$\bar K_z \equiv \bar k_{1z} + \bar k_{2z}$ with $\bar k_{iz}=\sqrt{E_i^2-\bar\varkappa_i^2}$.
	As we see, the asymmetry contrast remains very high and, as we scan over the total energy, it swings from almost $-1$ to $+1$.
	This is an unprecedented sensitivity to the photon polarization in a process which is fundamentally $P$-invariant.

\subsubsection{Helicity dependence: fixed-$\ell$ case}

Let us now adapt a different definition of what unpolarized twisted photons mean, 
see Section~\ref{subsection-unpolarized}.
When comparing $\gamma\gamma$ collisions with $\lambda_1 = \lambda_2 = \pm 1$,
we now assume that $m_1$ and $m_2$ change accordingly, so that 
$\ell_i \equiv m_i - \lambda_i$ are fixed. 
In this case, in the paraxial approximation, ${\cal J}_S$ becomes independent
of $\lambda$, see Eq.~\eqref{JS4}.
However, beyond the paraxial approximation, the difference persists.
Since the quantities ${\cal J}_1$ and ${\cal J}_2$ defined in Eq.~\eqref{JS3} now depend 
on $\lambda$, we replace \eqref{JS2} with
\be
{\cal J}_S = e^{i(\ell_1 - \ell_2)\varphi_{K}} \delta_{\lambda_1, \lambda_2} 
{2gE_1E_2 \varkappa_1 \varkappa_2 \over \Delta}\cdot \left({\cal J}'_1 + \lambda {\cal J}'_2\right)\,,
\ee
where
\bea
{\cal J}'_1 &=& \left[{(1+c_1)(1-c_2) \over 2} + {(1-c_1)(1+c_2)\over 2}\cos[2(\delta_1 +\delta_2)] 
- s_1s_2 \cos(\delta_1 +\delta_2) \right] \cos(\ell_1\delta_1 + \ell_2\delta_2)\,,\nonumber\\
{\cal J}'_2 &=& \left[s_1s_2 - (1-c_1)(1+c_2) \cos(\delta_1 +\delta_2) \right]
 \sin(\ell_1\delta_1 + \ell_2\delta_2) \sin(\delta_1 + \delta_2)\,.\label{JS5}
\eea
In the paraxial approximation $\theta_1 \ll 1$, $\bar\theta_2 \equiv \pi - \theta_2 \ll 1$,
\be
{\cal J}'_1 \to 2 \cos(\ell_1\delta_1 + \ell_2\delta_2)\,, \quad
{\cal J}'_2 \to \theta_1 \bar\theta_2  \sin(\ell_1\delta_1 + \ell_2\delta_2) \sin(\delta_1 + \delta_2) \ll {\cal J}'_1\,,
\ee
so that 
\be
A \approx \theta_1 \bar\theta_2  \tan(\ell_1\delta_1 + \ell_2\delta_2) \sin(\delta_1 + \delta_2) \ll 1\,.\label{asymmetry-contrast-2}
\ee
Thus, the non-zero asymmetry is suppressed by the small angles $\theta_1$, $\bar \theta_2$
but it may be additionally enhanced if a suitable position on the fringe is selected.

The above two evaluations of the polarization asymmetry of the twisted photon collision cross section
\eqref{asymmetry-contrast} and \eqref{asymmetry-contrast-2} differ significantly.
The real experimental situation will probably lie in between. 
Indeed, even if one produces twisted photons using holographic plates,
then one obtains, just behind the plate, a light field whose spacial distribution is not sensitive
to its polarization. However, the light field evolves downstream and will certainly be different in the focal plane
due to the intrinsic spin-orbital interaction of light \cite{beads2007}, as we discussed in section~\ref{subsection-unpolarized}.
Thus, the exact value of polarization asymmetry cannot be predicted without details of the experimental scheme.

However the mere fact of spontaneous generation of a (sizable) polarization asymmetry in twisted photon collisions
is beyond any doubt. This asymmetry is certainly absent in the usual plane photon-photon collision
and represents a novel experimental tool offered by twisted photons.

\subsection{Detecting scalar-pseudoscalar mixing in unpolarized twisted $\gamma\gamma$ collisions}

In the previous subsections, we demonstrated that unpolarized twisted photon collision
has a new intrinsic, adjustable degree of freedom, which is absent in the plane-wave case:
a difference between $\sigma_{\lambda = +1}$ and $\sigma_{\lambda = -1}$.
We will now show how it can be applied to detect scalar-pseudoscalar mixing
in a spin-0 resonance produced in collision of unpolarized twisted photons.
This is our first example of an observable which up to now was considered accessible
only in production of polarized photons or via the subsequent decays of the resonance produced.

Let us begin by considering production of a pseudoscalar particle $P$ in collision of two twisted photons.
The coupling is generated by the effective Lagrangian
\be
{\cal L}_P = i {g \over 4} F^{\mu\nu} \tilde F_{\mu\nu} P\,,
\ee
where $\tilde F_{\mu\nu} = \epsilon_{\mu\nu\rho\sigma} F^{\rho\sigma}/2$ is the dual electromagnetic field strength tensor.
It generates the following plane wave helicity amplitude
\be
{\cal M}_P = i g \epsilon_{\mu\nu\rho\sigma} k_1^\mu k_2^\nu e_1^\rho e_2^\sigma\,.\label{Mp} 
\ee
Working in the same Coulomb gauge, one can evaluate this amplitude explicitly 
to find the same structure as for the true scalar \eqref{Ms} times the overall helicity factor $\lambda$:
\be
{\cal M}_P = \lambda {\cal M}_S\,,
\ee
where for simplicity we used the same coupling constant $g$ in both cases.
For twisted photons, one concludes that ${\cal J}_P = \lambda {\cal J}_S$, which generates exactly the same cross section
as in the scalar case.
Thus, in the total production cross section, the pure scalar and pure pseudoscalar cases are as indistinguishable for twisted photon
collisions as for plane waves.

Next, suppose the spin-0 particle produced does not possess definite parity.
Then its production amplitude is
\be
{\cal M} = a {\cal M}_S + b {\cal M}_P = (a+\lambda b){\cal M}_S\,.
\ee
The (complex) coefficients $a$ and $b$ describe the scalar-pseudoscalar coupling of the particle to two photons.
In the usual plane wave collision with circularly polarized photons, the cross section is
\be
\sigma_\lambda \propto \left[|a|^2 + |b|^2 + 2\lambda \Re (a^* b)\right]\cdot |{\cal M}_S|^2\,.
\label{S-P-asym-PW}
\ee
By measuring the unpolarized production cross section, one can only reveal
the overall production intensity $|a|^2 + |b|^2$ but not detect the amount of scalar-pseudoscalar mixing.
It can be detected, in the plane wave collisions, only if one performs experiments with polarized photons 
and measures various spin asymmetries. For example, circularly polarized
photons give access to $\sigma_+ - \sigma_- \propto \Re (a^* b)$;
additional information can be recovered with linearly polarized photons.
This is a standard way to probe the parity properties of the produced resonance.

\begin{figure}[h]
	\centerline{
		\includegraphics[width=0.6\textwidth]{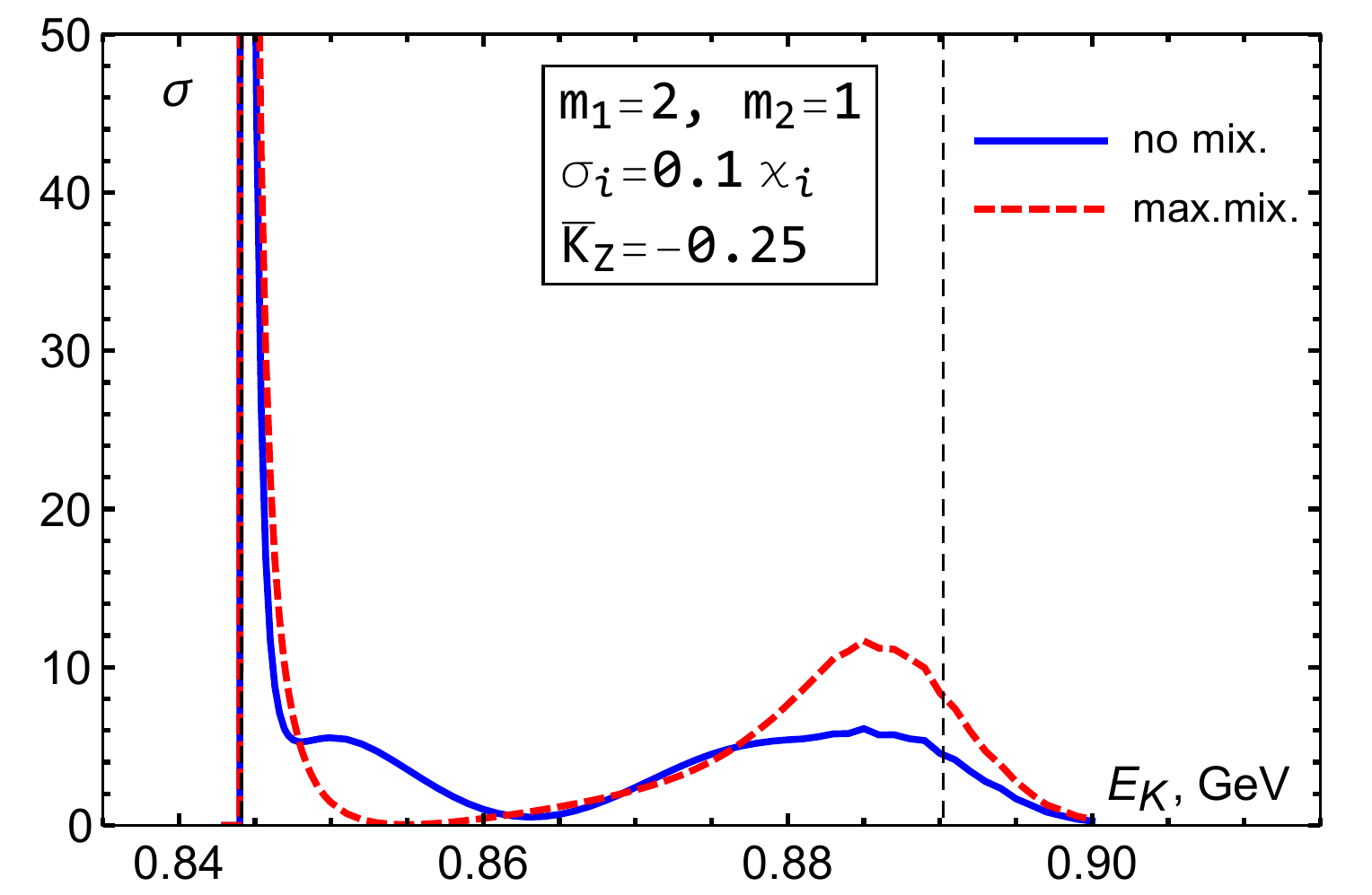}
	}
	\caption{
		\label{fig:SP-mixing}
		The total production cross section of a spin-0 state in twisted $\gamma\gamma$ collision with parameters \eqref{gamgan-parameters}
		with Gaussian-smeared twisted photons $\sigma_{i}=\varkappa_{i}/10$.
		Shown are cases without mixing (blue solid line) and with maximal scalar-pseudoscalar mixing (red dashed line).
	}
\end{figure}

Remarkably, twisted photons offer access to this scalar-pseudoscalar mixing
even with unpolarized twisted photons. 
Using the fixed-$m$ convention for unpolarized twisted light,
we obtain the twisted production amplitude as 
\be
{\cal J} = ({\cal J}_1 + \lambda {\cal J}_2)(a+ \lambda b) = (a {\cal J}_1 + b {\cal J}_2) + \lambda (b {\cal J}_1 + a {\cal J}_2)\,,
\ee
with helicity independent ${\cal J}_1$, ${\cal J}_2$ given in \eqref{JS3}.
Squaring it and averaging over the initial photon helicities, we obtain 
the unpolarized cross section as
\be
\sigma_+ + \sigma_- \propto ({\cal J}_1^2 + {\cal J}_2^2)(|a|^2 + |b|^2)
+ 4 {\cal J}_1{\cal J}_2\Re (a^* b)\,.\label{dsigma-S-P-tw}
\ee
Thus, even for unpolarized twisted photons, the cross section contains a term
which is sensitive to the magnitude of scalar-pseudoscalar mixing.

This contributions can be extracted by a scan of the total cross section over the collision energy.
Indeed, ${\cal J}_1$ and ${\cal J}_2$ in \eqref{JS3} have different dependence on $\delta_1$ and $\delta_2$.
Therefore, the ${\cal J}_1{\cal J}_2$ term exhibits a different interference pattern 
than ${\cal J}_1^2 + {\cal J}_2^2$ as one scans over the allowed energy interval.
To illustrate this effect, we present in Fig.~\ref{fig:SP-mixing} the total energy scan of the cross section
for a pure scalar or pseudoscalar (which are indistinguishable)
and for the case of their maximal mixing with $a = b = 1/\sqrt{2}$.
One sees a clear difference of the interference fringes in these two cases.
This particular plot corresponds to the kinematic parameters \eqref{gamgan-parameters}, 
but there certainly exists ample room for improving the discriminating power of this measurement.

\subsection{Spin physics with unpolarized twisted photons: $f_2$ example} 

Let us now see what unpolarized twisted photons can do for spin-2 resonances such as the $f_2(1270)$ meson.
In the usual plane wave case, unpolarized $\gamma\gamma$ collisions produce an equal amount of 
$\lambda_K$ and $-\lambda_K$ polarization states.
But in unpolarized {\em twisted} photon collisions, 
one can selectively produce different helicity states by adjusting the total collision energy 
in order to stay at an appropriate interference fringe. 

To see how it works, 
we begin by reviewing basic features of the $\gamma\gamma \to f_2$ process. 
The tensor meson $f_2$ with helicity $\lambda_K$ is described
with the symmetric polarization tensor $T_{\mu\nu}^{(\lambda_K)}$ orthogonal
to its four-momentum: $T_{\mu\nu}^{(\lambda_K)} K^\nu =0$.
It has five polarization states, with $\lambda_K$ spanning from $-2$ to $+2$.
They are constructed with the three polarization vectors $e^{(\lambda)}_\mu$, $\lambda = \pm 1, 0$, 
orthogonal to $K^\mu$: the vectors $e^{(\pm 1)}_\mu = (0, \bbe_{\bk \lambda})$
can be taken as defined in Eq.~\eqref{bbe-explicit} while 
$e^{(0)}_\mu = \gamma_K (\beta_K, \bn_K)$, where $\beta_K$ and $\gamma_K$
are the standard kinematic parameters for the produced meson and 
$\bn_K$ is the unit vector in the direction $\bK$.
The explicit expressions for the five polarization states of the spin-2 meson are
\bea
&&T_{\mu\nu}^{(\pm 2)} = e^{(\pm)}_\mu e^{(\pm)}_\nu\,, \quad 
T_{\mu\nu}^{(\pm 1)} = \fr{1}{\sqrt{2}}\left( e^{(\pm)}_\mu e^{(0)}_\nu + e^{(0)}_\mu e^{(\pm)}_\nu\right)\,,
\nonumber\\[2mm]
&&
T_{\mu\nu}^{(0)} = \fr{1}{\sqrt{6}}\left( e^{(+)}_\mu e^{(-)}_\nu + e^{(-)}_\mu e^{(+)}_\nu
+ 2 e^{(0)}_\mu e^{(0)}_\nu \right)\,.
\eea
The interaction between the two photons and the $f_2$-meson is generated by 
the Lagrangian $g F_{\rho\mu} F_{\rho\nu} T_{\mu\nu}/2$, which gives rise to the following plane wave 
$\gamma\gamma \to f_2$ amplitude:
\be
{\cal M} = g \left[(k_1 k_2) e_{1 \mu} e_{2\nu}
+ (e_1 e_2) k_{1 \mu} k_{2\nu}
- (k_1 e_2) e_{1 \mu} k_{2\nu}
- (k_2 e_1)  k_{1\mu}e_{2 \nu}
\right] (T_{\mu\nu}^{(\lambda_K)})^* \,.\label{f2-production}
\ee
Once again, all scalar products are understood as products of 4-vectors.
The polarization vectors for the two photons $e_1$ and $e_2$ depend on their helicities
$\lambda_1$ and $\lambda_2$ and are orthogonal to their respective momenta $k_1$ and $k_2$.
We work in the Coulomb gauge and use the same vectors \eqref{bbe-explicit}. 
As in the case of spin-0 production, we need the amplitude in the generic kinematics.
This is why we do not assume that $(k_1 e_2) = (e_1 k_2)= 0$ and keep all four terms in \eqref{f2-production}.

In the plane wave case, one can switch to the center of motion reference frame, in which
the produced $f_2$ is at rest and the photons are along the $z$ axis.
All polarization vectors in this case can be identified with the vectors ${\bm \chi}_{\lambda_i}$
defined in \eqref{chi}.
If one chooses the same axis $z$ to define the helicity of the $f_2$ meson, 
then the helicity amplitudes will take the following very simple form:
\be
{\cal M}_{\lambda_K = \pm 2} = 2 g E^2 \delta_{\lambda_1, \pm} \delta_{\lambda_2, \mp}\,.
\ee
That is, only $\pm 2$ polarization states can be produced and only for opposite photon helicities.
If the two photons have different energies $E_1 \not = E_2$
but their momenta are still along axis $z$, the final meson moves along the same axis
and has helicity $\pm 2$ depending on the initial photon helicities.
For unpolarized photon beams, the final $f_2$ meson is, of course, also unpolarized.

For unpolarized twisted photons, the produced $f_2$ meson has a non-vanishing average helicity.
To illustrate the main idea, we stick to the paraxial approximation for the two photons:
$\theta_1 \to 0$, $\theta_2 \to \pi$, while keeping the polar angle $\theta_K$ 
of the produced $f_2$ meson generic \eqref{theta-K}.
In the paraxial limit, the helicity amplitudes with $\lambda_K = \pm 2$ dominate and their expressions
take simple form. 
For $\lambda_K = +2$ we get
\bea
(\lambda_1,  \lambda_2 ) = (+, -): && {\cal M} = {g\over 2} E_1 E_2 (1+\cos\theta_K)^2 \,
e^{-i(\varphi_1 + \varphi_2 - 2\varphi_K)}\,,\nonumber\\
(\lambda_1,  \lambda_2 ) = (-, +): && {\cal M} = {g\over 2} E_1 E_2 (1-\cos\theta_K)^2 \,
e^{i(\varphi_1 + \varphi_2 - 2\varphi_K)}\,,
\eea
which gives ${\cal J}$, up to a common prefactor, of the form
\bea
(\lambda_1,  \lambda_2 ) = (+, -): && {\cal J} \propto 
(1+\cos\theta_K)^2 \cos(m_1\delta_1 + m_2\delta_2 - \delta_1 + \delta_2)\,,\nonumber\\
(\lambda_1,  \lambda_2 ) = (-, +): && {\cal J} \propto 
(1-\cos\theta_K)^2 \cos(m_1\delta_1 + m_2\delta_2 + \delta_1 - \delta_2)\,.
\eea
For $\lambda_K = -2$ we get
\bea
(\lambda_1,  \lambda_2 ) = (+, -): && {\cal J} \propto 
(1-\cos\theta_K)^2 \cos(m_1\delta_1 + m_2\delta_2 - \delta_1 + \delta_2)\,,\nonumber\\
(\lambda_1,  \lambda_2 ) = (-, +): && {\cal J} \propto 
(1+\cos\theta_K)^2 \cos(m_1\delta_1 + m_2\delta_2 + \delta_1 - \delta_2)\,.
\eea
It is immediately seen that for $\cos\theta_K = 0$
the two polarization states $\lambda_K = \pm 2$ are produced
in equal amount by the opposite photon helicities.
However at $\cos\theta_K\not = 0$, this equivalence breaks down.
Calculating $|{\cal J}|^2$ and averaging it over 
the initial photon helicities, we get the following unpolarized cross section
for $\lambda_K = \pm 2$:
\bea
\sigma_{\lambda_K = \pm2} &\propto& (1+6\cos^2\!\theta_K+\cos^4\!\theta_K) 
\left[1 + \cos(2m_1\delta_1 + 2m_2\delta_2)\cos(2\delta_1 - 2\delta_2)\right]\nonumber\\[1mm]
&&+ \ 2\lambda_K \cos\theta_K (1+\cos^2\!\theta_K)\cdot
\sin(2m_1\delta_1 + 2m_2\delta_2)\sin(2\delta_1 - 2\delta_2)\,.\label{f2J}
\eea
Thus, even if the twisted photons are unpolarized,
we do see a difference between $|{\cal J}_{+ 2}|^2$ and $|{\cal J}_{- 2}|^2$
and, therefore, between the production cross sections:
\be
\sigma_{\lambda_K = +2} - \sigma_{\lambda_K = -2} \propto
\cos\theta_K (1+\cos^2\theta_K)\cdot
\sin(2m_1\delta_1 + 2m_2\delta_2)\sin(2\delta_1 - 2\delta_2)\,.
\ee
For the longitudinally balanced collision, $k_{1z} + k_{2z} = 0$,
the emission angle is $\theta_K = \pi/2$, and this spin asymmetry vanishes.
But for an off-balanced situation with a generic $\theta_K$,
the asymmetry is present and, in general, not small.
Scanning the total energy of the collision and adjusting $m_i$, 
one can find optimal conditions when one produces preferentially $+2$ polarized states
over $-2$ states.

	We skip the numerical study of this effect for the $\gamma\gamma\to f_2$ production 
because in the following section we will study it at length for vector resonances
produced in the unpolarized twisted $e^+e^-$ annihilation.


\section{Spin asymmetry in twisted $e^+e^-$ annihilation}\label{section-e+e-}

Access to spin-dependent observables in unpolarized inclusive cross section
can also be expected from $e^+e^-$ annihilation, provided both the electron 
and the positron are twisted.
As discussed in Section~\ref{subsection-unpolarized},
an unpolarized twisted electron beam can be defined 
as an equal flux of twisted electrons with helicities $\zeta = +1/2$ and $\zeta = -1/2$ 
and either with fixed total angular momentum $m$ or fixed $\ell = m-\zeta$.
One can then calculate production of a vector meson with 
helicity $\lambda_K = \pm 1$ with unpolarized twisted electrons 
and observe a non-zero asymmetry $\sigma_{\lambda_K = +1} - \sigma_{\lambda_K = -1}$.
In this section, we will adopt the former definition of the unpolarized twisted electrons (fixed $m$), 
where a large effect is expected. What is actually feasible in experiment will
eventually depend on the exact scheme of preparation of twisted electrons and positrons.


Helicity amplitudes for vector (spin-1) meson production in
the plane wave annihilation process $e^-(k_1,\zeta_1)e^+(k_2,\zeta_2)\to V(K,\la_K)$ can defined as 
\begin{eqnarray}\label{eq:matr.el.ep.em.anih}
{\cal M}_{\zeta_1\zeta_2\la}=g \bar v_{\zeta_2}(k_2)\ga_\mu u_{\zeta_1}(k_1)V^{\mu *}_{\la_K}(K)\,,
\end{eqnarray}
Here for the sake of simplicity we assumed that the vector meson couples to the same $\bar v \gamma^\mu u$
current as the photon.
For realistic vector mesons, the current may differ according to whether the meson
represents an $S$-wave or $D$-wave state of the quark-antiquark pair \cite{Ivanov:1999pb},
but investigating this issue goes beyond the scope of the present paper.

The polarization vector $V^{\mu}_{\la_K}(k)$ is constructed in the same way as previously:
\be
V^{\mu}_{\pm 1}=(0,\bbe_{\bk, \pm 1})\,,\quad
V^{\mu}_{0}=\ga_K(\beta_K, \bn_K)\,,
\ee
where $\bbe_{\bk \lambda}$ is defined in Eq.~(\ref{bbe-explicit}).
 To simplify the calculations without losing the main features, 
we assume the electrons and positrons to be ultrarelativistic and neglect their mass.
Then one observes that the non-zero amplitudes exist
only for $\zeta_1 = -\zeta_2 \equiv \zeta$:
\be
{\cal M}_{\zeta_1\zeta_2\la_K} = - 2g \delta_{\zeta_1, -\zeta_2}\sqrt{E_1E_2}\cdot 
T_\mu^{(\zeta)} V^{\mu *}_{\la_K}\,,
\ee
where
\begin{eqnarray}
T_0^{(\zeta)} &=& {w_2^{(\zeta)}}^\dagger w_1^{(\zeta)} = 
c_1c_2 e^{i\zeta(\varphi_2-\varphi_1)} + s_1s_2 e^{-i\zeta(\varphi_2-\varphi_1)}\,,\nonumber\\
T_3^{(\zeta)} &=& 2\zeta\, {w_2^{(\zeta)}}^\dagger \sigma_3 w_1^{(\zeta)} = 
c_1c_2 e^{i\zeta(\varphi_2-\varphi_1)} - s_1s_2 e^{-i\zeta(\varphi_2-\varphi_1)}\,,\nonumber\\
T_1^{(\zeta)} &=&  2\zeta\, {w_2^{(\zeta)}}^\dagger \sigma_1 w_1^{(\zeta)} = 
c_2s_1 e^{i\zeta(\varphi_2+\varphi_1)} + s_2c_1 e^{-i\zeta(\varphi_2+\varphi_1)}\,,\nonumber\\
T_2^{(\zeta)} &=&  2\zeta\, {w_2^{(\zeta)}}^\dagger \sigma_2 w_1^{(\zeta)} = 
-2i\zeta \left[c_2s_1 e^{i\zeta(\varphi_2+\varphi_1)} - s_2c_1 e^{-i\zeta(\varphi_2+\varphi_1)}\right]
\,.\label{Tmu}
\end{eqnarray}
One can verify that $T_\mu^{(\zeta)} k_1^\mu = T_\mu^{(\zeta)} k_2^\mu = 0$.
We remind the reader that, for fermions, the shorthand notation is to be understood 
as $c_i \equiv \cos(\theta_i/2)$ and $s_i \equiv \sin(\theta_i/2)$.

These helicity amplitudes can be evaluated for generic kinematics, 
but the main features again can be illustrated in the paraxial approximation:
$\theta_1 \to 0$, $\theta_2 \to \pi$, so that 
only terms with $c_1 \to 1$ and $s_2 \to 1$ survive.
The polar angle $\theta_K$ of the produced resonance is kept generic.
The surviving helicity amplitudes correspond to production of $\lambda_K = \pm 1$ states:
\be
{\cal M}_{\zeta, -\zeta, \la_K} = - g \sqrt{2E_1E_2} e^{-i\zeta(\varphi_2+\varphi_1 - 2\varphi_K)}
\cdot (\lambda_K\cos\theta_K + 2\zeta) \,.\label{Me+e-}
\ee
When passing from plane waves to twisted Bessel states,
we use the same expression for ${\cal J}$ as in Eq.~\eqref{J2}, with half-integer $m_1, m_2$. 
The plane wave amplitudes ${\cal M}_a$, ${\cal M}_b$ 
for the two kinematic configurations are given by Eq.~\eqref{Me+e-},
in which the azimuthal angles take their values set by Eqs.~\eqref{phi12}.
The resulting expression for the twisted amplitude is
\be
{\cal J}_{\zeta, -\zeta, \la_K}= - g\sqrt{2E_1E_2}
e^{i(m_1-m_2)\varphi_K}\frac{\varkappa_1\varkappa_2}{\Delta}
\cdot (\lambda_K\cos\theta_K + 2\zeta)\, 
\cos[m_1\delta_1 + m_2\delta_2 - \zeta(\delta_1 - \delta_2)]\,.\label{Je+e-}
\ee
Using the fixed-$m$ definition of the unpolarized electron and positron beams,
we obtain the following unpolarized twisted $e^+ e^-$ cross section in the paraxial approximation:
\begin{eqnarray}\label{eq:cross.sec.paraxial}
\sigma_{\la_K = \pm 1}&\propto& 
(1+\cos^2\theta_K)[1+\cos(2m_1\de_1+2m_2\de_2) \cos(\de_1-\de_2)] \\
&& +\  2\lambda_K \cos\theta_K \sin(2m_1\de_1+2m_2\de_2)\sin(\de_1-\de_2)\,.
\end{eqnarray}
This results is similar to the $f_2$-meson production in unpolarized twisted photon collision:
for a generic $\theta_K \not = \pi/2$, there is a clear imbalance
between $\lambda_K = +1$ and $\lambda_K = -1$ production cross sections,
which depends on the exact position with respect to the interference fringes.
The asymmetry contrast can be defined as
\be
\label{eq:asymmetry.exact}
A = \frac{\si_{\lambda_K = +1}-\si_{\lambda_K = -1}}{\si_{\lambda_K = +1}+\si_{\lambda_K = -1} + \si_{\lambda_K = 0}}\,,
 \ee
and in the paraxial limit it can be approximated as 
\begin{eqnarray}\label{eq:asymmetry.paraxial}
A \approx
\frac{2\cos\theta_K}{1+\cos^2\theta_K} \cdot \frac{\sin(2m_1\de_1+2m_2\de_2)\sin(\de_1-\de_2)}
{1+\cos(2m_1\de_1+2m_2\de_2) \cos(\de_1-\de_2)}\,.
\end{eqnarray}
Thus, for the production angles $\theta_K$ far away from $\pi/2$,
the asymmetry can be rather sizable.
Notice that in the paraxial limit $\sigma_{\lambda_K=0}\to 0$.

These paraxial estimates can be corroborated with numerical calculations
based on the exact expression for the helicity amplitudes \eqref{Tmu}.
To give a concrete example, we consider production of the $J/\psi$ meson 
with mass $M=3.1$~GeV 
in the unpolarized twisted $e^+e^-$ annihilation with the following parameters:
\begin{eqnarray}
\varkappa_1=0.2~\text{GeV}\,,\quad \varkappa_2=0.1~\text{GeV}\,, \quad m_1=5/2\,,\quad  m_2=1/2\,.\label{eq:nonsmearedparameters}
\end{eqnarray}
To define the scan trajectory on the $(E_1, E_2)$ plane, we will first choose the electron energy $E_1$
and then scan over a range of the positron energies $E_2$ plotting the cross section
as a function of the final particle polar angle 
$\theta_K$ calculated from Eq.~(\ref{theta-K}).

\begin{figure}[h]
	\centerline{
		\includegraphics[width=0.5\textwidth]{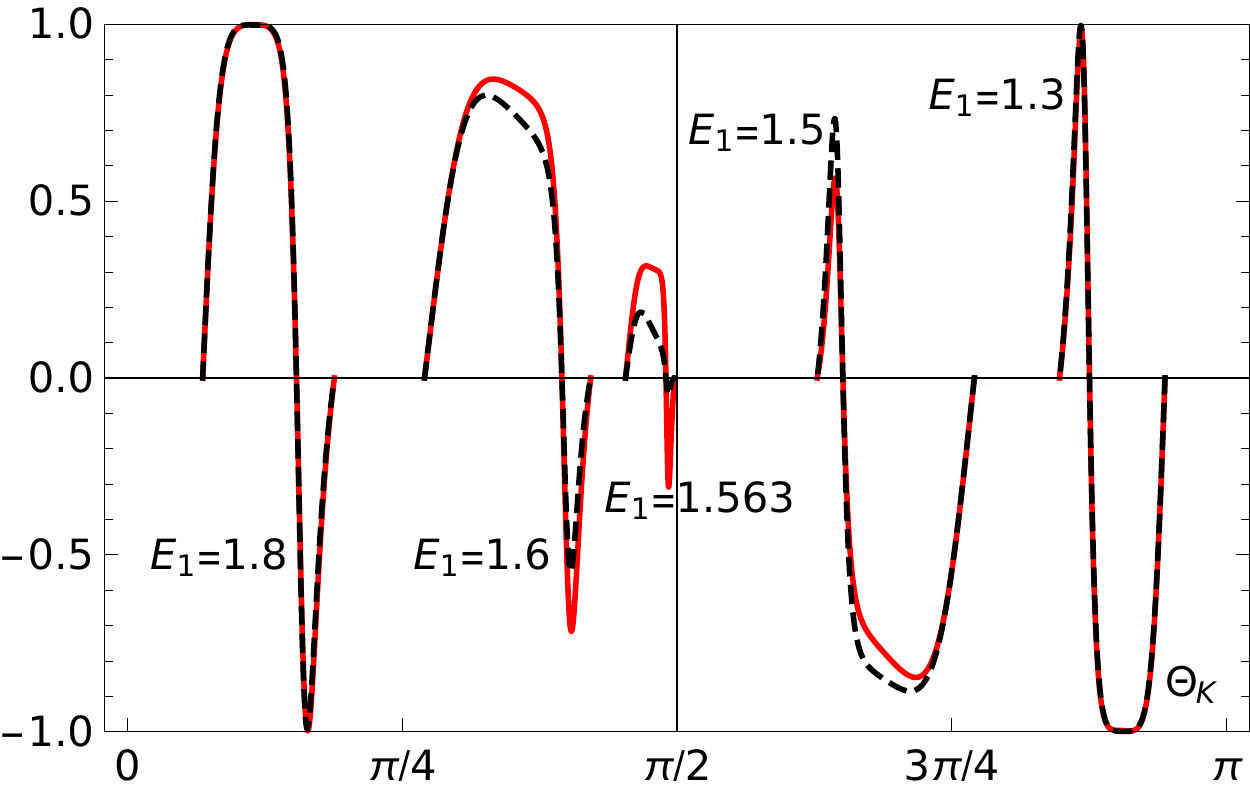}
	}
	\caption{
		\label{fig:asymmetry-e1-variation}
		The polarization asymmetry $A$ in Eq.~(\ref{eq:asymmetry.exact})
		as a function of the final particle polar angle $\theta_K$ 
		with the parameters given in \eqref{eq:nonsmearedparameters} and
		for several choices of the electron energy $E_1$ given in GeV.
		The black dashed lines are the asymmetries in
		paraxial limit, while the red solid curves represent the exact result 
		without including the $\sigma_{\lambda_K=0}$ contribution to Eq.~(\ref{eq:asymmetry.exact}).
	}
\end{figure}

In Fig.~\ref{fig:asymmetry-e1-variation}
we plot the polarization asymmetry for pure Bessel beams
with different values of the electron energy $E_1$ ranging from 1.3 GeV to 1.8 GeV
and with $E_2$ scanned in a certain range for each $E_1$.
We see that the exact numerical results are very well approximated by the paraxial
limit for the kinematic configuration away from $\theta_K = \pi/2$.
The sign of the predominant polarization asymmetry is clearly correlated with
the forward or backward production hemisphere.
For a fixed energy $E_1$, the curve is non-symmetric,
which implies that even if electron and positron beams have certain energy spread,
the overall polarization asymmetry will persist.

We stress that the distributions shown in Fig.~\ref{fig:asymmetry-e1-variation}
are not to be understood as the angular distribution in a {\em single} fixed-energy experiment.
A single experiment with fixed energies $E_1$ and $E_2$ will correspond to one specific angle $\theta_K$
with its cross section and polarization.
The plots in Fig.~\ref{fig:asymmetry-e1-variation} represent 
the evolution of the polar production angle and the correlated evolution 
of the spin asymmetry value as one scans over the initial positron energy, keeping the electron energy fixed. 
They tell us that selecting a point near the plateau would produce almost $100\%$ polarized meson beam
even with unpolarized initial electron and positron beams.

Since the exact Bessel beams are not normalizable,
we model the realistic situation by smearing over the initial $\varkappa_i$ with the following parameters:
\begin{eqnarray}\label{eq:smearingParameters}
\bar\varkappa_1=0.2~\text{GeV}\,,~
\bar\varkappa_2=0.1~\text{GeV}\,,~
\si_i=\bar\varkappa_i/5\,,~
E_1=1.8~\text{GeV}\,,~
E_2=1.338~\text{GeV}\,,
\end{eqnarray}
and for the angular momentum values $m_1=5/2$ and $m_2=1/2$.
Although we now fix the energies of both incoming particles, smearing over $\varkappa_i$
produces a distribution over a range of angles $\theta_K$ in a single experiment.
We also take into account the finite width of the produced resonance and 
evaluate the differential cross section weighted with the corresponding Breit-Wigner factor
with the width $\Gamma=93$~keV:
\begin{eqnarray}\label{eq:BWeffects-s}
\frac{d\sigma}{ds\,d\cos\theta_K}&\sim&  
\sqrt{E_K^2-s} ~|\lr{{\cal J}}|^2
\frac{1}{\pi}\frac{M\Gamma}{(s-M^2)^2+M^2\Gamma^2}\,,
\\\label{eq:BWeffects}
\frac{d\sigma}{d\cos\theta_K}&\sim&
\frac{1}{\sqrt{E_K^2-M^2}}
\int\limits_0^{E_K}ds\, \frac{d\sigma}{ds\,d\cos\theta_K}\,,
\end{eqnarray}
where $s\equiv K_\mu K^\mu =E_K^2-K_z^2-K^2$ is the four-momentum squared of the final particle. 
We stress that, unlike in the plane wave collision, the variable $s$ is not fixed by the initial state kinematics.
Even when $E_i$ and $\varkappa_i$ are fixed, resonances with masses within a certain interval can be produced \cite{Ivanov:2019pdt}.
For resonances with finite width, this intrinsic mass-spectrometric feature of the twisted particle collision manifests itself
as an $s$-distribution of the cross section.
The dependence of the matrix element on the final particle parameters comes through $s$ and the polar angle $\theta_K$.

\begin{figure}[t]
	\centerline{
	\includegraphics[width=0.49\textwidth]{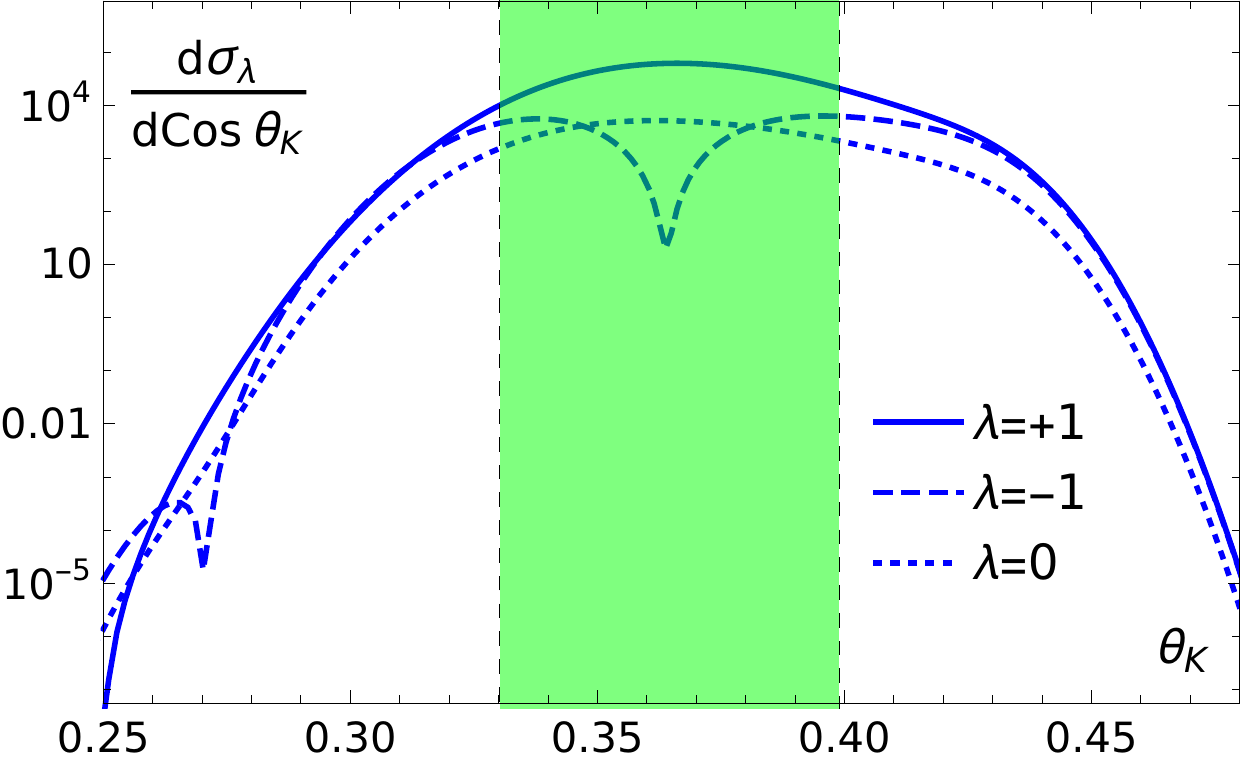}\hspace{5mm}
	\includegraphics[width=0.47\textwidth]{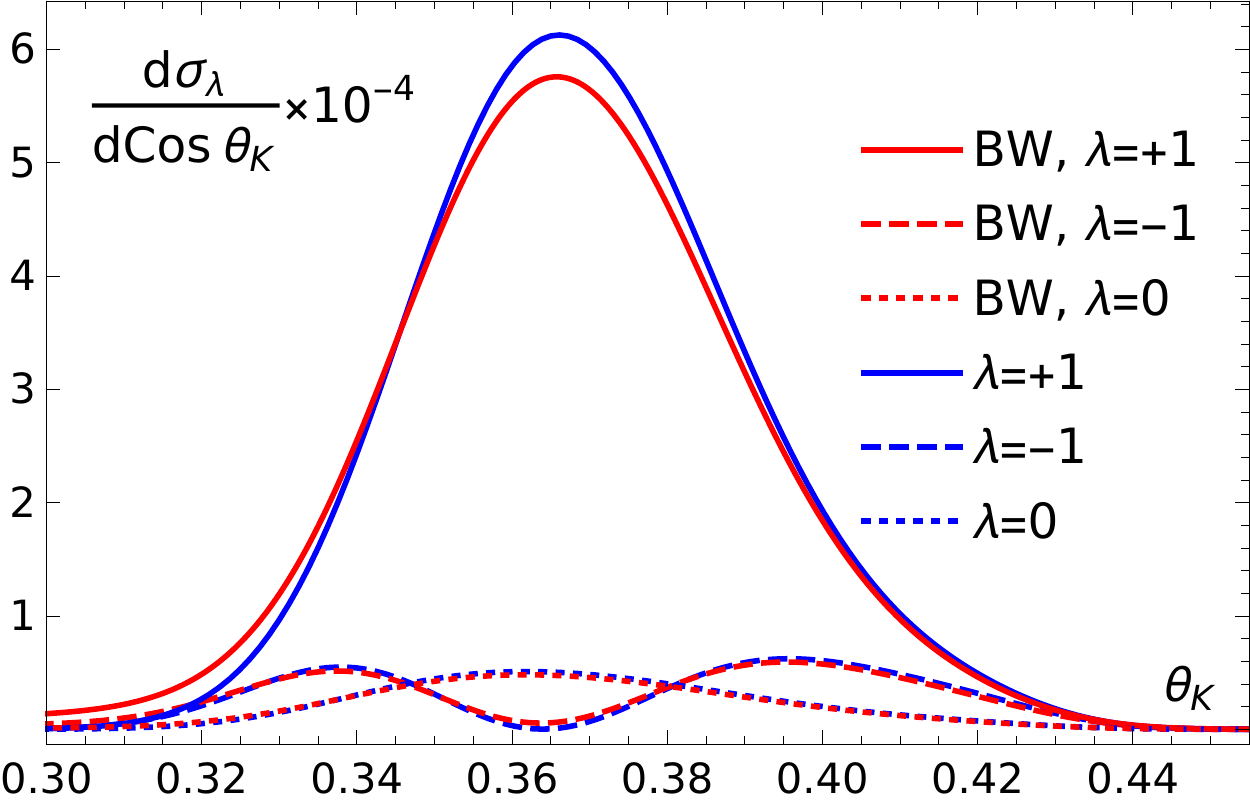}
}
	\caption{
		\label{fig:crossX4ePeM-annihilation}
	Left: Distribution over the polar angles $\theta_K$ of the production of the $J/\psi$ meson in the zero-width approximation
	in the unpolarized twisted $e^+ e^-$ annihilation with kinematic parameters as in Eq.~(\ref{eq:smearingParameters}).
    Solid, dashed, and short-dashed lines show the cross sections for $\lambda_K = +1$, $-1$, and $0$, respectively.
	Right: the same cross sections in linear scale. The red curves include finite width effects according to Eq.~(\ref{eq:BWeffects}),
	while the blue curves correspond to the zero width. 
	The green band denotes the angular interval, which gives the main contribution to the cross section.
	}
\end{figure}

In Fig.~\ref{fig:crossX4ePeM-annihilation} we show the resulting differential cross sections $d\si_{\lambda_K}/d\cos\theta_K$ 
for all three polarization states $\la_K=\pm 1, 0$.
The left plot shows these cross sections with zero width in the log scale,
while the right plot, presenting the same functions in the linear scale,
illustrates the minor effect of the non-zero width.
The green band indicates the angular range which saturates the cross section.

\begin{figure}[h!]
	\centerline{
		\includegraphics[width=0.49\textwidth]{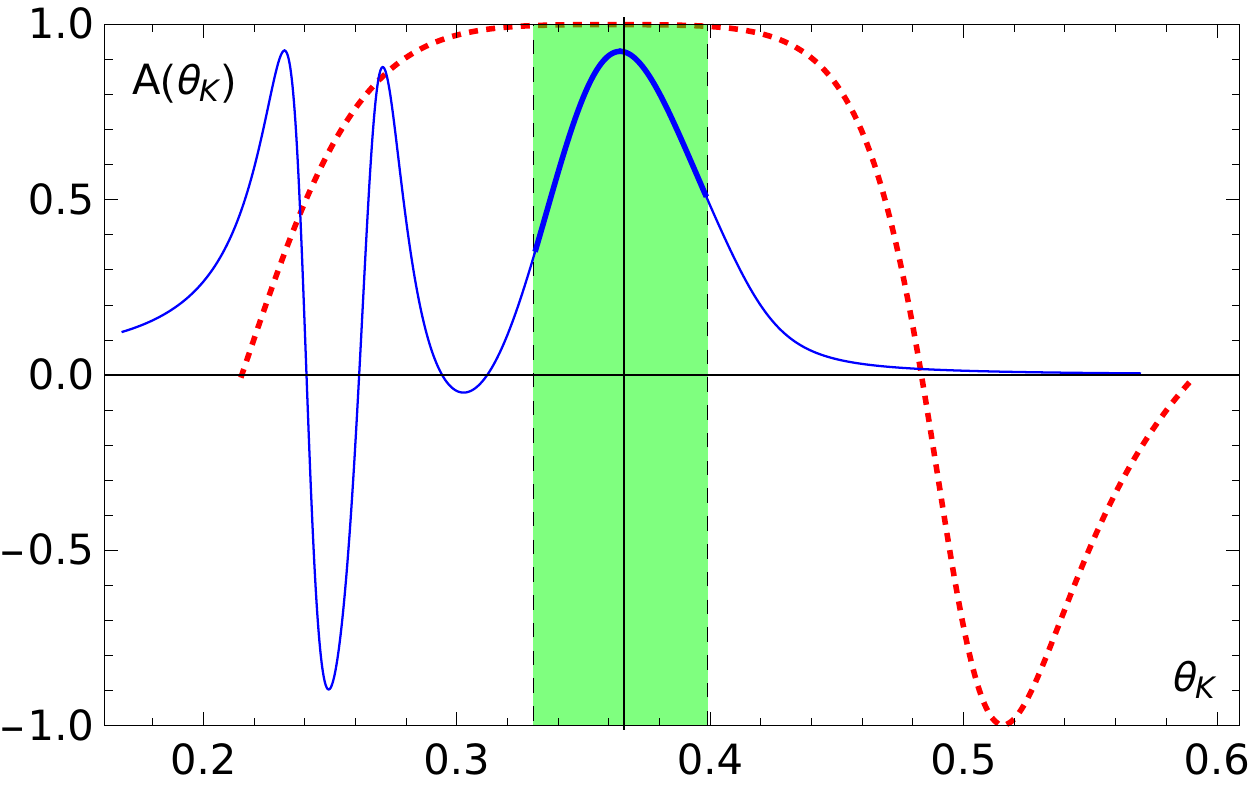}			
	}
	\caption{
		\label{fig:asymmetry4ePeM-smearing}
		The differential polarization asymmetry $A$, Eq.~(\ref{eq:asymmetry-diff.exact}),
		as a function of the polar angle $\theta_K$ of the produced resonance. 
		The red dotted line represent the result for the pure Bessel beams 
		and is identical to the leftmost curve in Fig.~\ref{fig:asymmetry-e1-variation}.
		The blue solid line represents the asymmetry with smearing effects taken into account
		and with fixed energies of initial particles.
		The green band is the same as in Fig.~\ref{fig:crossX4ePeM-annihilation}, left. 
		The vertical line around $\theta_K\approx 0.36$ represent the angle with
			$K_z = \bar k_{1z} + \bar k_{2z}$, where $\bar k_{iz} = \sqrt{E_i^2 - \bar\varkappa_i^2}$. 
	}
\end{figure}

As one sees, the cross section around the peak is strongly dominated by the polarization state 
$\lambda_K = +1$, with a $\approx 10\%$ admixture of the $\lambda_K = 0$ state
and even smaller contribution from $\lambda_K = -1$.
This is not a coincidence but is a result of our choice of the kinematic parameters \eqref{eq:smearingParameters}. 
Certainly, by adjusting these parameters, one can arrange for a situation with $\lambda_K = -1$ production
strongly dominating over $\lambda_K = +1$.
Thus, we obtain a remarkable result: we can produce almost fully polarized vector mesons
in unpolarized twisted $e^+e^-$ annihilation.

The polarization purity of the produced resonance
can be quantified by the differential asymmetry $A(\theta_K)$ defined as
\begin{equation}
\label{eq:asymmetry-diff.exact}
A(\theta_K) = \left({d\si_{\lambda_K = +1}\over d\cos\theta_K}-{d\si_{\lambda_K = -1} \over d\cos\theta_K}\right)\biggr/
{\sum_{\la_K} {d\si_{\la_K} \over d\cos\theta_K}}\,.
\end{equation}
This quantity is plotted in Fig.~\ref{fig:asymmetry4ePeM-smearing}.
Although the smearing effects reshape the angular profile of the asymmetry,
it still rises as high as $90\%$.

\begin{figure}[h]
	\centerline{
		\includegraphics[width=0.47\textwidth]{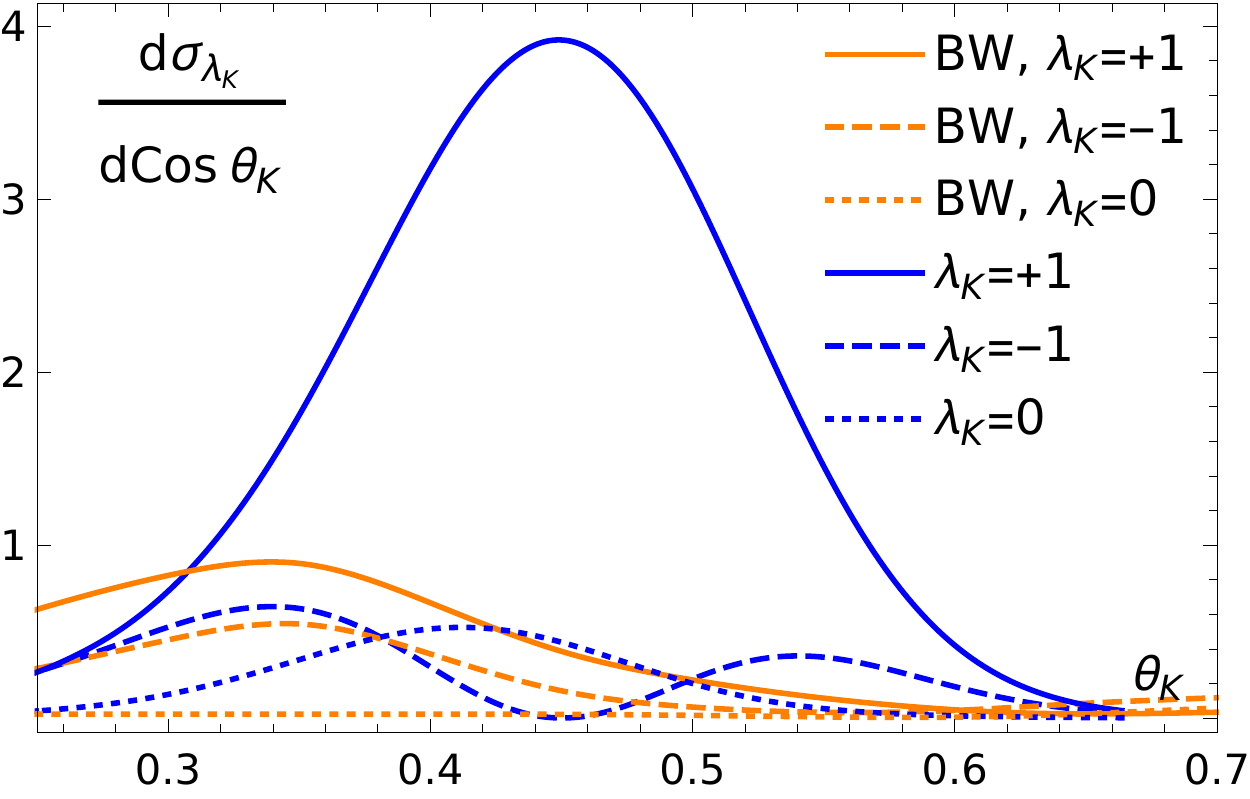}
	}
	\caption{
		\label{fig:asymmetry-rho}
		The cross sections (in arbitrary units) as in Fig.~\ref{fig:crossX4ePeM-annihilation}, right, but for the wide $\rho$-meson.
		The lighter orange curves include finite width effects according to Eq.~(\ref{eq:BWeffects}),
		while the blue curves correspond to the zero width.
	}
\end{figure}

\begin{figure}[h!]
	\centerline{
		\includegraphics[width=0.49\textwidth]{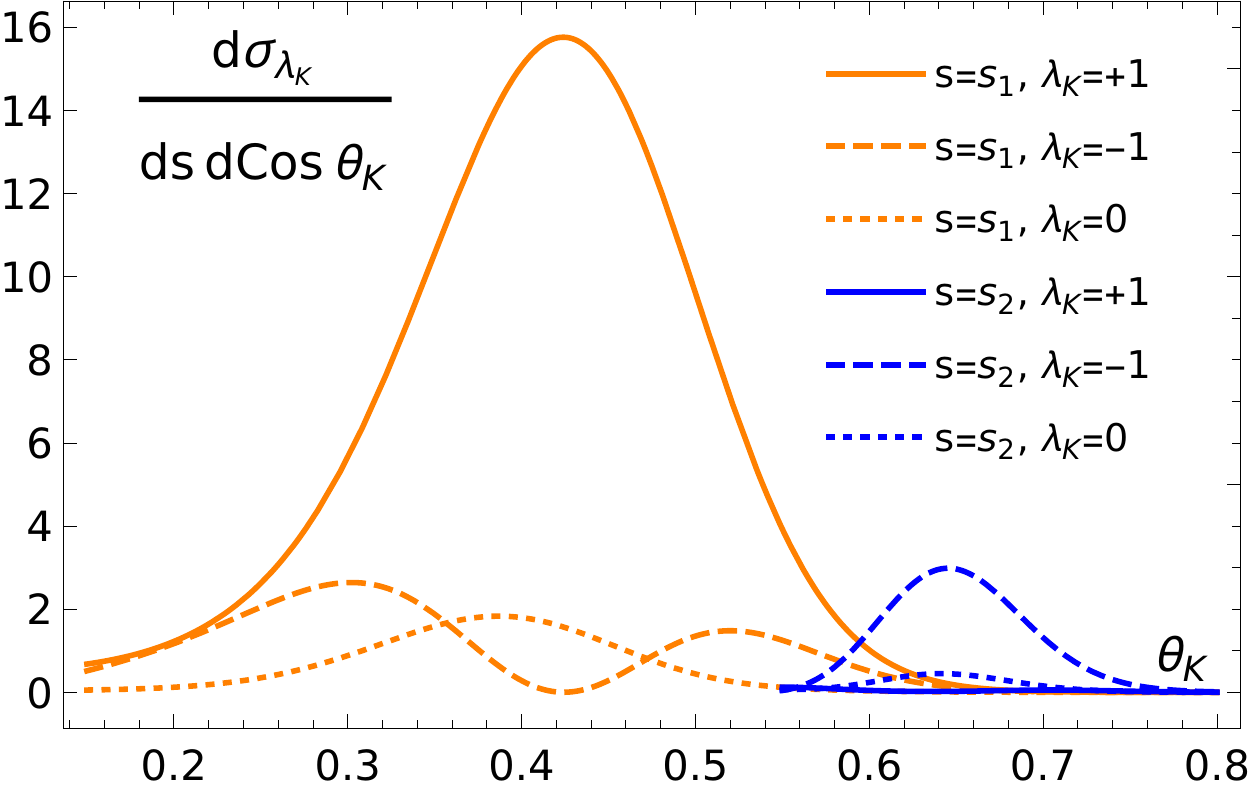}\hspace{5mm}
		\includegraphics[width=0.49\textwidth]{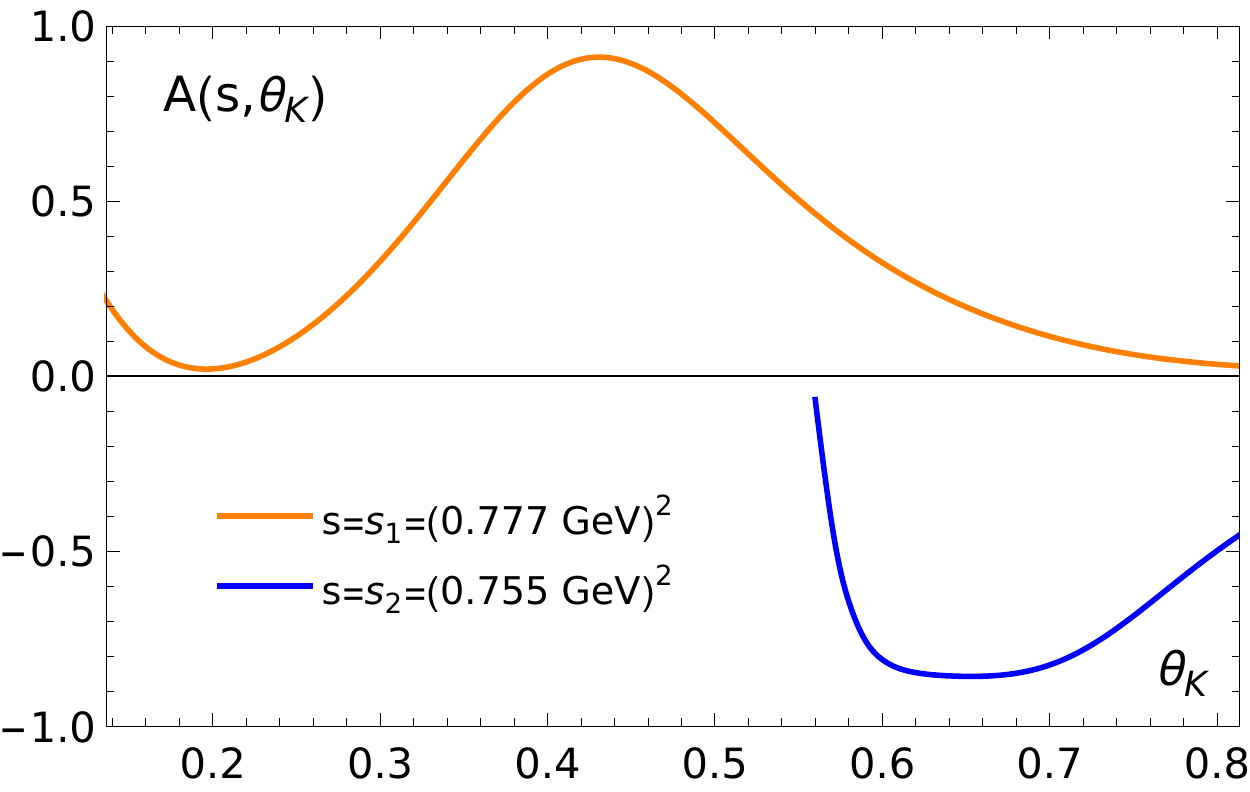}
	}
	\caption{
		\label{fig:crossX4ePeM-annihilation-rhomeson}
		Polarization properties of $\rho$ mesons produced in the unpolarized twisted $e^+ e^-$ annihilation 
		with kinematic parameters as in Eq.~(\ref{eq:smearingParameters-rho}).
		Left: Differential cross section (in arbitrary units) for the broad range of polar angles $0.1 < \theta_K < 0.8$
		and for the two values of the center of mass energies: 
		$\sqrt{s_1}=0.777$~GeV (lighter orange curves) and $\sqrt{s_2}=0.755$~GeV (darker blue curves). 
		The solid, dashed, and short-dashed curves show the cross sections for $\lambda_K = +1$, $-1$, and $0$, respectively.
		Right: Differential polarization asymmetry $A$, Eq.~(\ref{eq:asymmetry-diff.exact}),
		as a function of the polar angle $\theta_K$. 
		The finite width effects are included according to Eq.~(\ref{eq:BWeffects-s}).
	}
\end{figure}

In the above numerical example, we used a very narrow resonance. 
Let us now see how the picture changes if one considers
a wide resonance such as $\rho$ meson 
with mass $M_\rho =0.775$~GeV and width $\Gamma_\rho=0.149$~GeV.
We take the angular momentum values $m_1=5/2$ and $m_2=1/2$
and the following kinematical parameters:
\begin{eqnarray}\label{eq:smearingParameters-rho}
\bar\varkappa_1=0.2~\text{GeV}\,,~
\bar\varkappa_2=0.1~\text{GeV}\,,~
\si_i=\bar\varkappa_i/5\,,~
E_1=0.6~\text{GeV}\,,~
E_2=0.258~\text{GeV}\,.
\end{eqnarray}
In Fig.~\ref{fig:asymmetry-rho} we show the angular distribution of the $\rho$ production cross section
in unpolarized twisted $e^+e^-$ annihilation for the three helicities of the produced meson.
The dramatic broadening effect of the large width is evident. 
Nevertheless, the cross sections with $\lambda_K = +1$ and $-1$ differ significantly,
so that the value of the asymmetry is non-zero and rather large.

The effect becomes even more pronounced if one studies the same angular distribution
for selected values of $\sqrt{s}$. We remind the reader that even when initial kinematical parameters 
$E_i$ and $\varkappa_i$ are fixed, the invariant mass of the produced resonance is not fixed and can vary 
within certain range \cite{Ivanov:2019pdt}. 
However, since $\rho$ decays to the $\pi\pi$ system, the detector can reconstruct 
the invariant mass and the direction of the produced $\rho$ meson,
enabling us to plot the fixed $\sqrt{s} = M_{\pi\pi}$ slice of the angular distribution.

The resulting differential cross sections $d\si_{\lambda_K}/(ds\,d\cos\theta_K)$ 
are presented in Fig.~\ref{fig:crossX4ePeM-annihilation-rhomeson} 
together with differential polarization asymmetry for two values of 
invariant mass 	$\sqrt{s}=0.777$~GeV and $\sqrt{s}=0.755$~GeV.
The left plot shows these cross sections for all three polarization states $\la_K=\pm 1, 0$,
while the right plot demonstrates the polarization asymmetry.
The curves clearly show that, at fixed initial energy of the $e^+ e^-$ collision,
the polarization of the producted meson dramatically depends on the production angle.
For example, at $\sqrt{s}=0.777$~GeV (light red curves),
the asymmetry $A$ reaches the values of $90\%$ around $\theta_K = 0.42$.
By slightly lowering the collision energy to $\sqrt{s}=0.755$~GeV (dark blue curves),
one completely reverses the situation: now the polarization state $\la_K=-1$ dominates,
and the asymmetry reaches $-85\%$ in the $\theta_K$ region $0.6$ to $0.7$. 
These broad plots clearly show that obtaining highly polarized $\rho$-mesons
does not require any fine-tuning nor very narrow angular selection.
Self-polarization is an intrinsic, robust feature of this production scheme.

\section{Discussion and conclusions}\label{section-discussion}

Uncovering and exploring the spin properties of hadrons and their interactions
is an intricate and fascinating topic in hadron phenomenology.
There is a wealth of information revealed through spin-dependent observables
in hadronic processes but it is not always easy to extract and disentangle them experimentally.
Although the spin physics program pursued by the hadronic community
at specialized colliders is rich and multifaceted, 
any new complementary method of accessing spin observables would be very welcome.

In this work we explored in detail the idea that we recently proposed in \cite{Ivanov:2019vxe} 
that spin- and parity-dependent observables can be studied even with unpolarized particles,
provided they are prepared in {\em twisted} states.
In such a state, particles possess a well-defined $z$-projection of the angular momentum,
which receives contribution not only from spin but also from the orbital angular momentum.
If one prepares a beam of photons or electrons with equal amount of positive and negative helicities
but with the angular momentum fixed, then the resonance production cross sections
will display dramatic energy dependence and angular effects, which will reveal the spin-dependent
observables in a novel way.

We showed how production of a hypothetical spin-0 particle in collision of unpolarized twisted photons
can reveal the amount of its scalar-pseudoscalar mixing.
This is an illustration of the power of twisted particles in detecting parity-violating effects
in the fully unpolarized case.
We also demonstrated that vector mesons produced in unpolarized twisted $e^+e^-$ annihilation 
can in fact be almost $100\%$ polarized and their polarization state can be controlled
by adjusting kinematics of the colliding twisted particles.
None of these effects is possible with the usual plane wave collisions.

All these opportunities offer a remarkably rich pattern of observable spin physics effects 
which can be probed in twisted particle collisions, although not yet at existing colliders. 
Twisted electrons and photons have been experimentally demonstrated 
only for low energies, and one needs first 
to prepare high-energy twisted particles and bring them into collisions.
This field is barely explored, but there exist theoretical suggestions such as \cite{Jentschura:2010ap,Jentschura:2011ih} 
which await exploration.
We believe that the novel opportunities in hadronic physics offered by twisted particles
present a sufficiently compelling scientific case to justify further dedicated work
on their realization.

\section*{Acknowledgments}
I.P.I. thanks the Institute of Modern Physics, Lanzhou, China, for financial support and hospitality during his stay.
I.P.I. acknowledges funding from the Portuguese
\textit{Fun\-da\-\c{c}\~{a}o para a Ci\^{e}ncia e a Tecnologia} (FCT) through the FCT Investigator 
Contract IF/00989/2014/CP1214/CT0004 under the IF2014 program,
and through Contracts PTDC/FIS-PAR/29436/2017, UID/FIS/00777/2019, and CERN/FIS-PAR/0004/2017,
which are partially funded through Programa Operacional Competitividade e Internacionaliza\c{c}\~{a}o (POCI), 
Quadro de Refer\^{e}ncia Estrat\'{e}gica Nacional (QREN), and the European Union.
I.P.I. also acknowledges the support from National Science Center, Poland,
via the project Harmonia (UMO-2015/18/M/ST2/00518).
P.M.Z. and N.K. are supported by the National Natural Science Foundation of China 
via Grants No. 11975320 (P.M.Z.) and No. 11875296 (N.K.).
A.V.P. and N.K. thank the Chinese Academy of Sciences President's International Fellowship Initiative for the support
via Grants No. 2019PM0036 (A.V.P.) and No. 2017PM0043 (N.K.).
The work has been partially supported by the Ministry of Education and Science of the Russian Federation: 
Projects No. 3.6371.2017/8.9 and No. 3.6439.2017/8.9.


\end{document}